\documentclass[%
 twocolumn,
 amsmath,amssymb,
 aps,
]{revtex4-1}

\usepackage{graphicx}
\usepackage{dcolumn}
\usepackage{bm}
\usepackage{subcaption}
\usepackage[parfill]{parskip}
\usepackage{epstopdf}
\usepackage{float}
\usepackage[justification=raggedright]{caption}
\usepackage{tikz}
\usetikzlibrary{arrows}
\usepackage{xcolor}

\newcommand\crule[3][black]{\textcolor{#1}{\rule{#2}{#3}}}

\definecolor{dred}{HTML}{AE2D2D}

\begin{document}

\title{Segmented ion-trap fabrication using high precision stacked wafers}
\author{Simon Ragg}%
 \altaffiliation[]{Contributed equally.}
\author{Chiara Decaroli}
 \altaffiliation[]{Contributed equally.}
 \author{Thomas Lutz}
\author{Jonathan P. Home}%
\affiliation{%
 Trapped Ion Quantum Information Group, Institute for Quantum Electronic,\\ ETH Zurich, 8093 Zurich,\\ Switzerland}

\date{\today}

\begin{abstract}
We describe the use of laser-enhanced etching of fused silica in order to build multi-layer ion traps. This technique offers high precision of both machining and alignment of adjacent wafers. As examples of designs taking advantage of this possibility, we describe traps for realizing two key elements of scaling trapped ion systems. The first is a trap for a cavity-QED interface between single ions and photons, in which the fabrication allows shapes that provide good electro-static shielding of the ion from charge build-up on the mirror surfaces. The second incorporates two X-junctions allowing two-dimensional shuttling of ions. Here we are able to investigate designs which explore a trade-off between pseudo-potential barriers and confinement at the junction center. In both cases we illustrate the design constraints arising from the fabrication.

\end{abstract}

\maketitle

\section{INTRODUCTION}\label{Sec:Intro}

Atomic ions trapped in radio frequency traps cooled and controlled by laser light provide an experimental platform which is extremely well isolated from environmental effects. As a result, this setting is among the leading candidates for quantum computing \cite{bruzewicz2019, Brown2011single,Harty2014high,Benhelm2008towards,Ballance2016prl} as well as playing an important role in precision measurements and frequency standards \cite{highlychargedions, Pietisotope, atomicclocksnist}. In both areas extending the current levels of control to larger numbers of ions is important. This is obvious for a quantum computer, which will rely on manipulating a large number of qubits. For atomic clocks an increase in the number of ions under control would provide improved signal-to-noise ratio \cite{Herschbach2012}, while including techniques known from quantum computing could further enhance this through the use of entanglement \cite{Blatt2008}.

One challenge of putting together systems with more ions is the realization of suitable trap structures which can be easily and repeatably fabricated. These must meet a number of requirements, including having precise and uniform structures, allowing good optical access, and being able to withstand the high voltages and resulting electric fields (including radio-frequency fields) which are commonly used. As these systems become more extensive, the alignment of the electrode structures becomes increasingly important \cite{2014highprecisiontrap}. In quantum computing, particular challenges arise for trap development in the context of the ``Quantum CCD'' architecture, in which ions are dynamically shuttled through multiple connected zones of a trap array during the execution of algorithms \cite{Kielpinski2002architecture,Blakestad2009high}. For scaling it may also be necessary to interface the ion with photonic connections between remote modules. Two elements which are integral to these approaches are junction regions which can guide ions in two dimensions, and single-ion/single-photon interfaces. In the long term scaling may also require the integration of a variety of components into the trap structure, such as optics for light delivery and collection \cite{Eltony2012,VanRynbach2016}.

A number of approaches have been taken previously in an attempt to realize suitable traps using techniques which can potentially be scaled to larger systems while maintaining precision. Common requirements are that electrode structures of a few 100 micron size can be realized, while retaining a precision of the electrode boundaries at close to 1 micron. One important area of research involves traps which are produced by lithography \cite{Charles_Doret_2012,Seidelin2006,Pearson2005}, which allows the production of monolithic structures which are very precisely defined (at the level of a few 10s of nanometers). However it is difficult in many of these processes to fabricate multi-layer stacks to thicknesses of more than a few tens of microns \cite{Stick2006}. The heating of ions near conducting surfaces decreases rapidly with distance, thus ion-electrode distances used for experiments have all been $> 30$~$\mu$m, which is greater than the lithographic thickness. Because of this most lithographic traps have been surface-electrode traps, in which the ions are trapped above a single plane of electrodes, providing a strong asymmetry in the out-of-plane potential. This configuration results in a sacrifice of trap depth and curvature for fixed ion-electrode distance and voltage compared to more symmetric electrode designs. This makes tasks such as shuttling through two-dimensional junctions more challenging \cite{Blakestad2009high, Wright2013}.

Micro-fabricated ion traps with heights of 100 micron or more have been produced using either stacks of wafers which are aligned manually \cite{Kienzler2015, BlakestadNIST2010}, or in a monolithic fashion using silicon fabrication techniques developed for Micro-Electrical-Mechanical-Systems (MEMS) \cite{Brownnutt2006, WilpersNature2012}. Wafer stacking techniques have generally relied on combinations of optical monitoring and manual alignment, which is limited to tens of microns. Several experiments using stacked laser-machined alumina wafers have reported suspected misalignment as a cause of undesirable intrinsic micromotion \cite{gaebler2016, negnevisky2018}. This reduces the interaction strength between an ion and a single-frequency laser beam, thus impeding control, and gives rise to an important systematic shift in frequency standards \cite{Mehlstaubler2012}. The monolithic approach potentially overcomes this problem \cite{Brownnutt2006,WilpersNature2012}, but these traps have not become reliably available to experimentalists and thus are relatively unexplored.

In this article, we describe the use of laser-enhanced etching to construct precisely machined 3-dimensional ion traps based on stacked fused silica wafers. We use the precision and flexibility of this manufacturing method to design self aligning stackable structures, detail the capabilities and constraints of this approach, and give guidance in designing traps. We present two trap designs for quantum computing experiments which take advantage of these methods. The first is a segmented linear trap for which the electrode structures are designed to provide a high level of electro-static shielding of the ion from potential charge build up on an externally mounted optical cavity. The second is a segmented trap featuring two junctions, which should allow flexible re-configuration of arrays of ions. We give the results of testing the wafer alignment, as well as detailing the full fabrication chain that we have developed.

\section{Femtosecond laser-enhanced etching}\label{Sec:FemtosecondLaserEnhancedEtching}

\begin{figure}[t]
\centering
\includegraphics[width=0.35\textwidth]{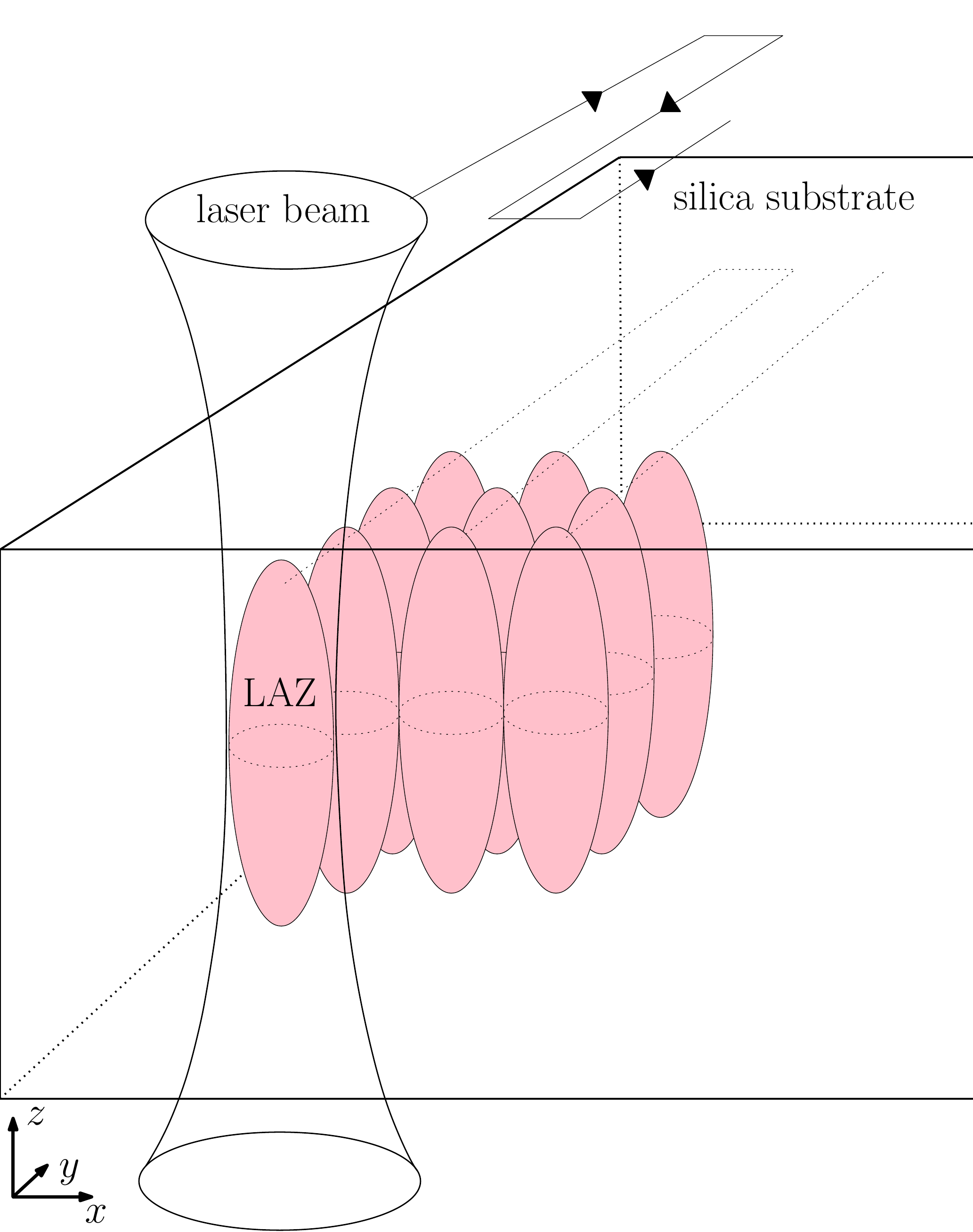}
\put(180,0){\textcolor{white}{t}}
\caption{Shape and arrangement of the volume modified by single laser pulses \cite{BellouardMRS2004}, which is known as Laser Affected Zone (LAZ).}
\label{Fig:LaserImpactZoneFemtoMachin}
\end{figure}

Femtosecond laser-enhanced etching consists of exposing a glass substrate to femtosecond laser light followed by wet-etching with a dilute hydrofluoric acid solution \cite{MarcinkeviciusOL2001}. It has been observed that silica exposed to femtosecond laser light has a significantly faster etching rate compared to not exposed silica \cite{KondoJJAP1998,MarcinkeviciusOL2001}. This is believed to be due to the creation of an internal stress field and a change in the chemical structure of the silica \cite{BellouardOSA2004}.
Due to the non-linear nature of femtosecond laser interaction with silica, enhancement of the etching rate occurs only in a small ellipsoidal volume in the focus of the beam referred to as the Laser Activated Zone (LAZ) \cite{BellouardMRS2004}, shown in Figure \ref{Fig:LaserImpactZoneFemtoMachin}. Thus, the full volume to be etched needs to be written into the substrate by the laser. It has also been observed that in volumes with multiple laser exposures the etching rate is even higher than in domains that have been exposed only once \cite{BellouardOSA2004}.

In recent years several experiments and applications have demonstrated the strength of this technology. Applications include laser written optical waveguides \cite{BellouardMRS2004,DavisOL1996,StreltsovJOSA2002,HiraoNCS1998,MiuraAIP1997}, creation of high aspect ratio micro-fluidic channels \cite{BellouardOSA2004,BellouardMRS2004,BellouardOME2011}, flextures \cite{BellouardOME2011}, MEMS \cite{LenssenAPL2012} and ion traps \cite{YoshimuraEPJ2015,An2018,Dugan2015}. We are currently aware of two companies offering a customizable femtosecond laser machining service: Translume \cite{TranslumeWeb} in the United States, and FEMTOprint \cite{FEMTOprintWeb,FemtoprintProject} in Switzerland. The work below has been carried out with devices fabricated by the latter.

\subsection{Capabilities of the machining process}\label{Sec:CapabilitiesMachiningProcess}

In the following sections we will discuss the strengths and main limitations of femtosecond laser machining of silica with regard to ion trap fabrication. We focus on the achievable tolerances stated by FEMTOprint and give an overview of our own experiences with test devices designed while working with FEMTOprint's team.
We focus on the machining precision, the achievable aspect-ratio of 3-dimensional structures such as grooves and cuts and the ability to create small yet precisely defined 3-dimensional geometries.

As a result of the shape of the LAZ, the specified process precision of femtosecond laser-enhanced etching is $\pm\ 2\ \mu\mathrm{m}$ in the direction of propagation of the laser beam (hereafter these will be referred to as horizontal surfaces, with the laser beam assumed to propagate vertically) and $\pm\ 1\ \mu\mathrm{m}$ \cite{FEMTOprintWeb} orthogonal to it (these surfaces we refer to as vertical). On our test samples we have observed that these values represent a ``best case scenario", which can be achieved in localized regions. In larger structures and over full wafers we observe machining tolerances on the 10 micron level. 

\onecolumngrid

\begin{figure}[H]
\begin{subfigure}{\textwidth}
  \includegraphics[width=.5\textwidth]{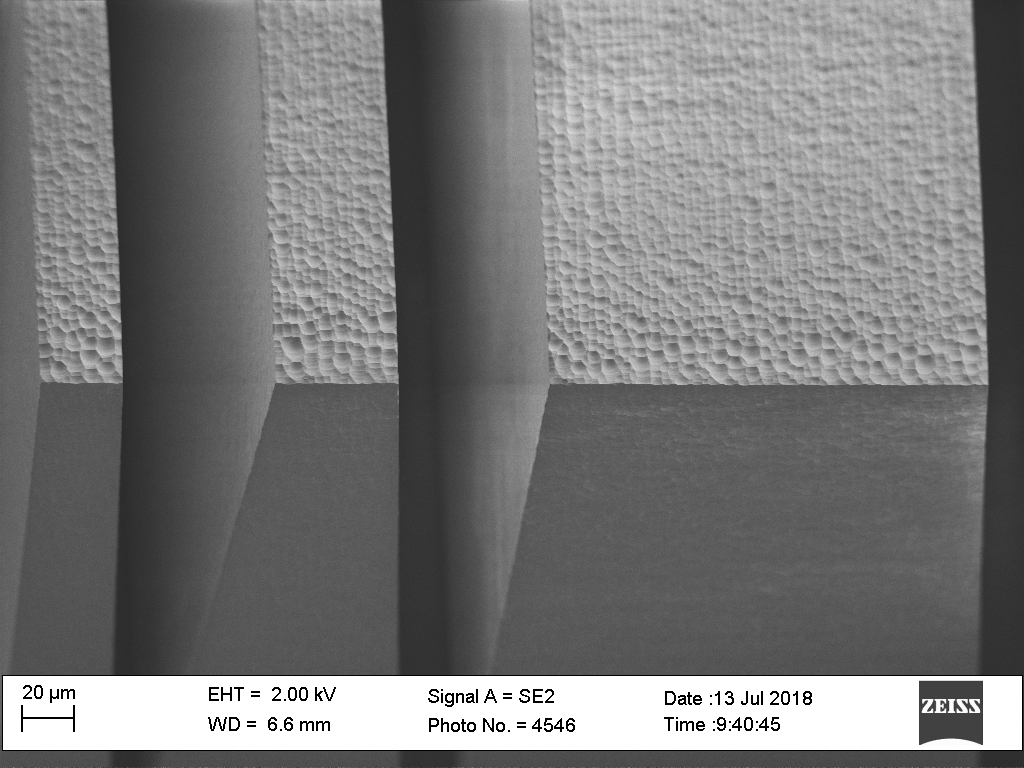}
  \includegraphics[width=.5\textwidth]{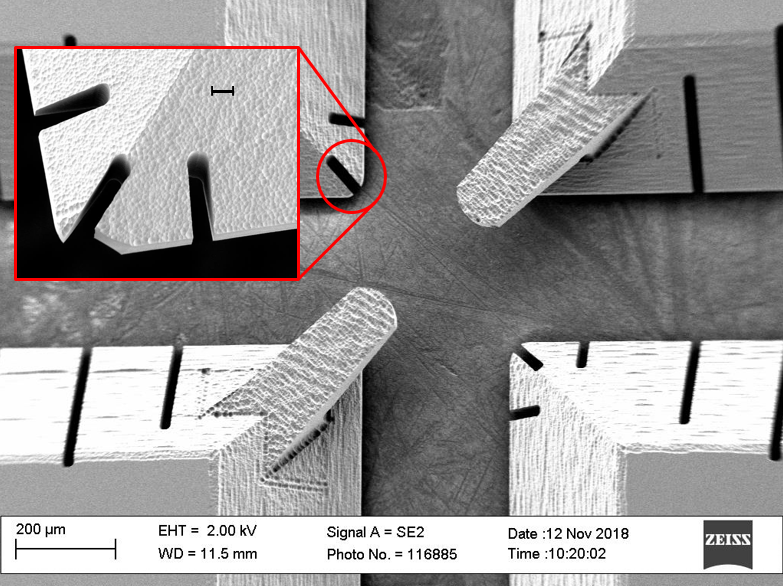}
  \put(-450,148){\includegraphics[width=.37\textwidth]{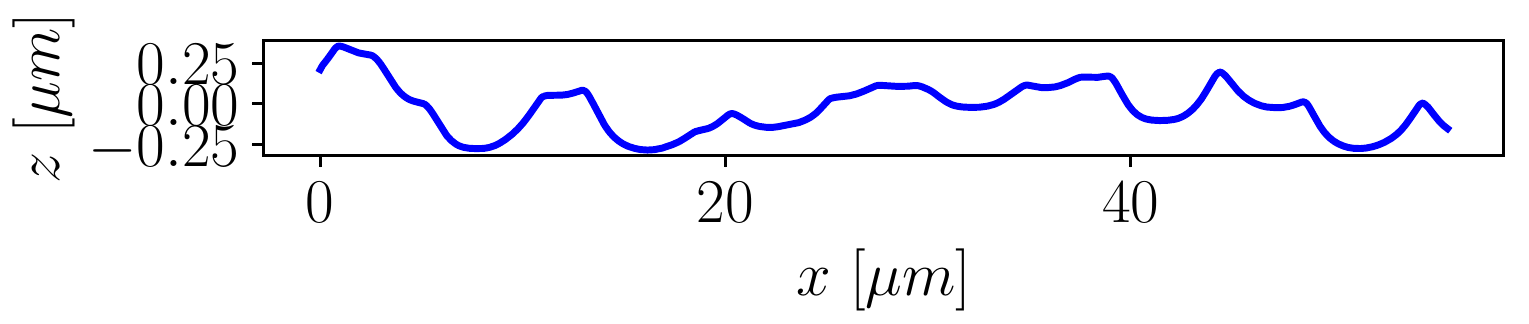}}
  \put(-350,134){
  \begin{tikzpicture}
    \draw[blue,line width=0.6mm] (0,0) -- (1.2,0);
  \end{tikzpicture}}
  \put(-453,135){
  \begin{tikzpicture}
    \draw[blue,thick] (0,0) -- (-3.65,0.45);
  \end{tikzpicture}}
  \put(-316,135){
  \begin{tikzpicture}
    \draw[blue,thick] (0,0) -- (1.8,0.45);
  \end{tikzpicture}}
  \put(-316,105){\footnotesize{\textcolor{white}{\textbf{horizontal}}}}
  \put(-316,80){\footnotesize{\textcolor{white}{\textbf{vertical}}}}
  \put(-509,177){\crule[white]{.50cm}{.4cm}}
  \put(-507,180){\footnotesize{\textcolor{black}{\textbf{a.)}}}}
  \put(-252,177){\crule[white]{.50cm}{.4cm}}
  \put(-250,180){\footnotesize{\textcolor{black}{\textbf{b.)}}}}
  \put(-193,167){\footnotesize{\textbf{20 $\mu$m}}}
    \put(-85,103){
  \begin{tikzpicture}[thick]
    \draw [white,   -latex      ] (0,0.0) -- (-0.7,0.3) node [right] {};
  \end{tikzpicture}
  }
  \put(-60,100){\small{\textcolor{white}{protrusion}}}
  \label{fig:TopSurfaceFingers}
\end{subfigure}
\begin{subfigure}{\textwidth}
\includegraphics[width=.249\textwidth,trim={9.cm 0cm 9.cm 0cm},clip]{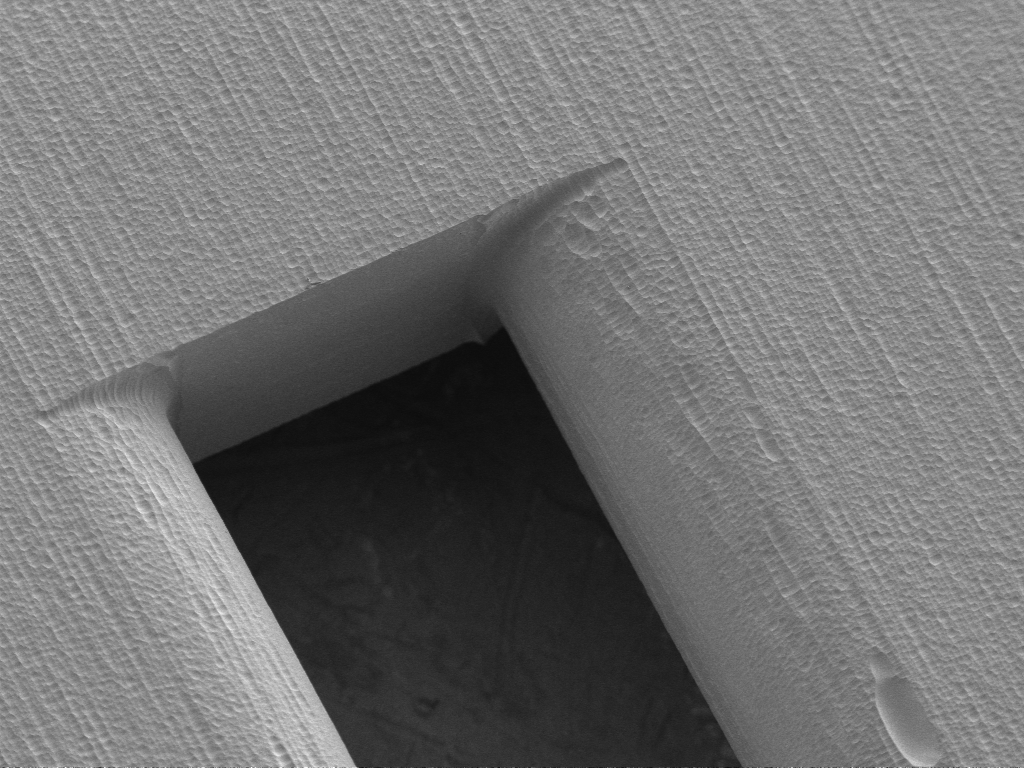}
\includegraphics[width=.249\textwidth,trim={9.cm 0cm 9.cm 0cm},clip]{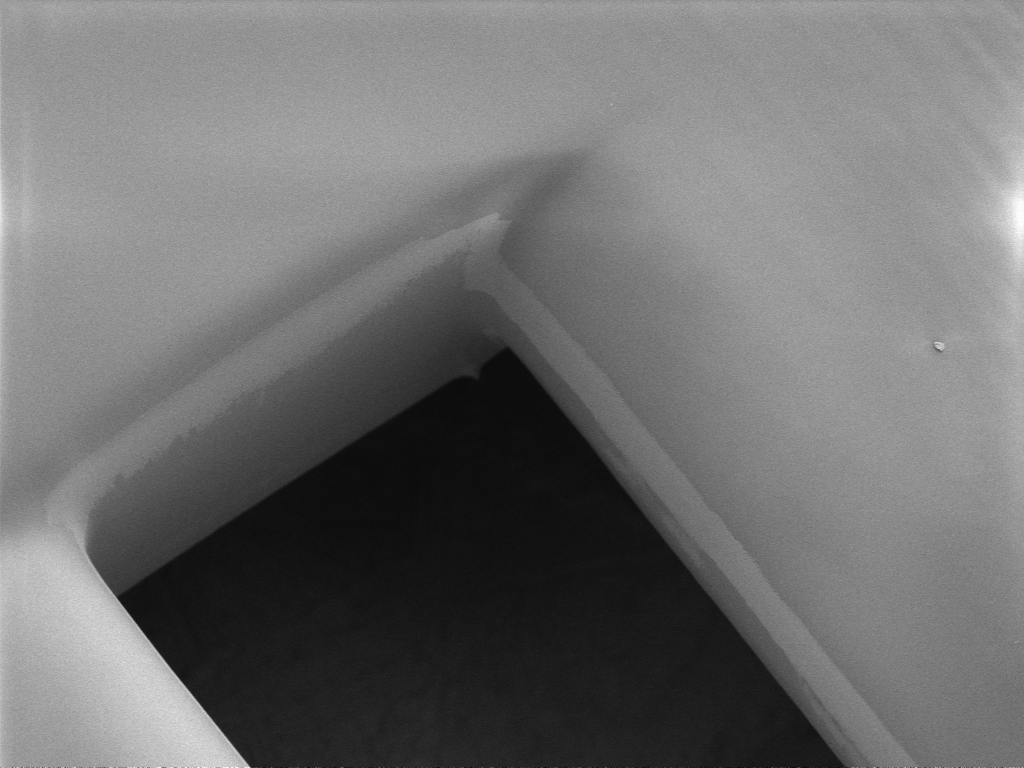}
  \includegraphics[width=.5\textwidth]{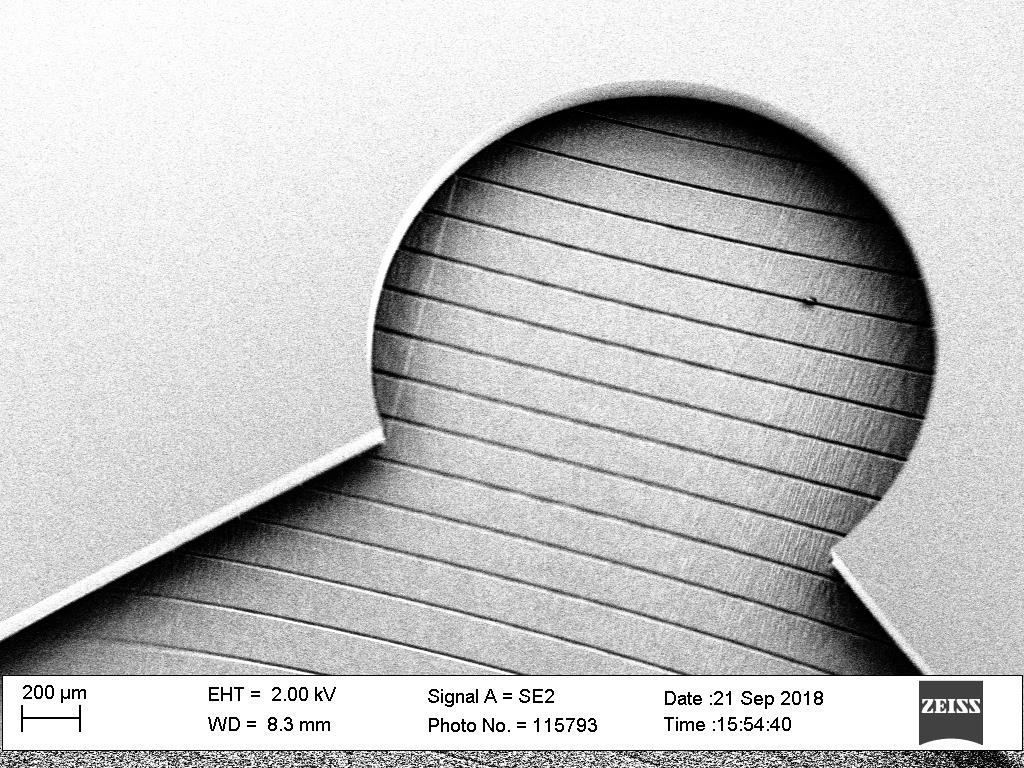}
  \put(-175,103){
  \begin{tikzpicture}[thick]
    \draw [red,   -latex      ] (0,0.6) -- (1,0.0) node [right] {};
  \end{tikzpicture}
  }
    \put(-250,125){\footnotesize{\textcolor{red}{\textbf{undesired grooves}}}}
  \put(-512,4){\includegraphics[width=.5\textwidth,trim={0.cm 0.55cm 0.cm 23.8cm},clip]{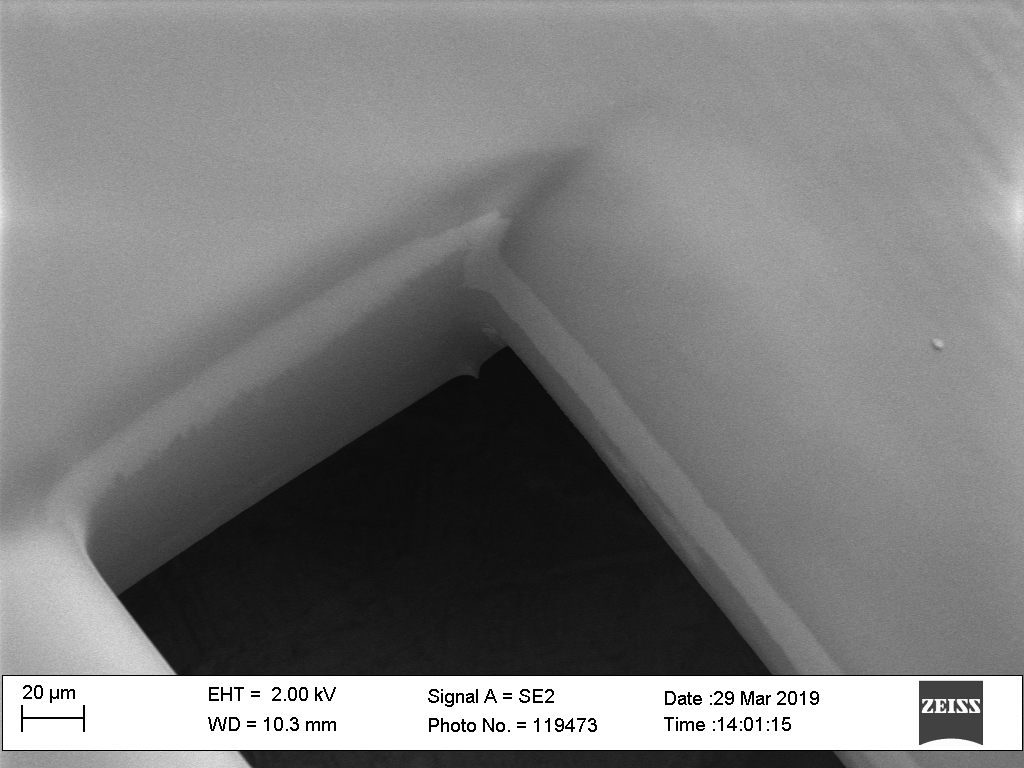}}
    \put(-480,30){\includegraphics[width=.37\textwidth]{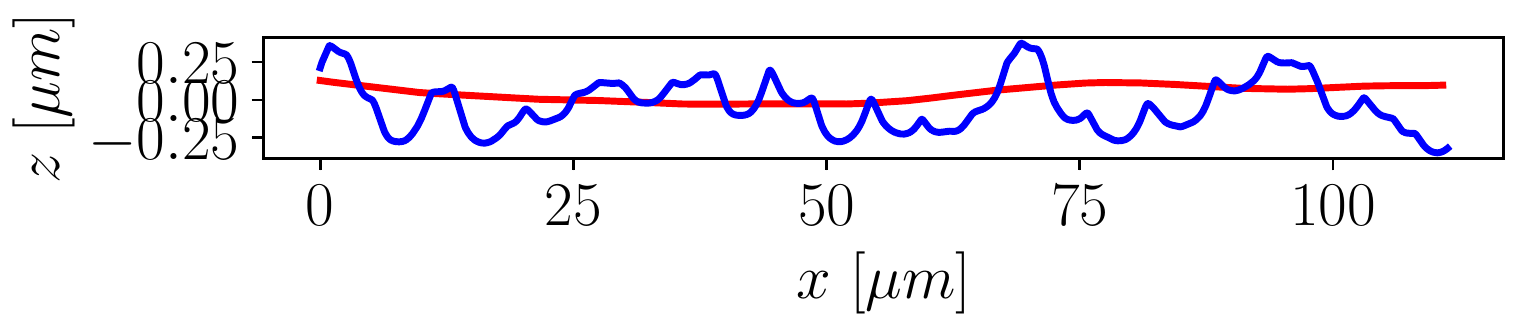}}
    \put(-509,177){\crule[white]{.50cm}{.4cm}}
  \put(-507,180){\footnotesize{\textcolor{black}{\textbf{c.)}}}}
  \put(-252,177){\crule[white]{.50cm}{.4cm}}
  \put(-250,180){\footnotesize{\textcolor{black}{\textbf{d.)}}}}
  \put(-360,180){\footnotesize{\textcolor{red}{\textbf{after polishing}}}}
  \put(-480,180){\footnotesize{\textcolor{blue}{\textbf{before polishing}}}}
  \label{fig:TopSurfaceFingers}
\end{subfigure}
\caption{\textbf{a.)} SEM image of a silica sample with segmented electrodes  separated by tapered gaps. The horizontal surface features the characteristic surface structure of femtosecond laser machining while the vertical side walls have less roughness. The inset shows the surface profile along the horizontal surface. \textbf{b.)} Test junction trap with small free standing 3D structure as well as angled surfaces and small electrodes. The inset shows a common defect, a dent on a sharp corner, which may be due to high stress in these small areas during etching. \textbf{c.)} On the left: a sample of a machined surface. On the right: the same surface after laser polishing. Both SEM images have been taken with the same settings. The inset compares the surface profiles of machined surfaces before (blue) and after (red) polishing. \textbf{d.)} Defects (deep machining grooves) on a machined, horizontal surface. Large amount of material have been removed through intense laser exposure. The grooves are in the order of 10 $\mu$m wide and deep.}
\label{fig:SEMImageFingerTestSample}
\end{figure}
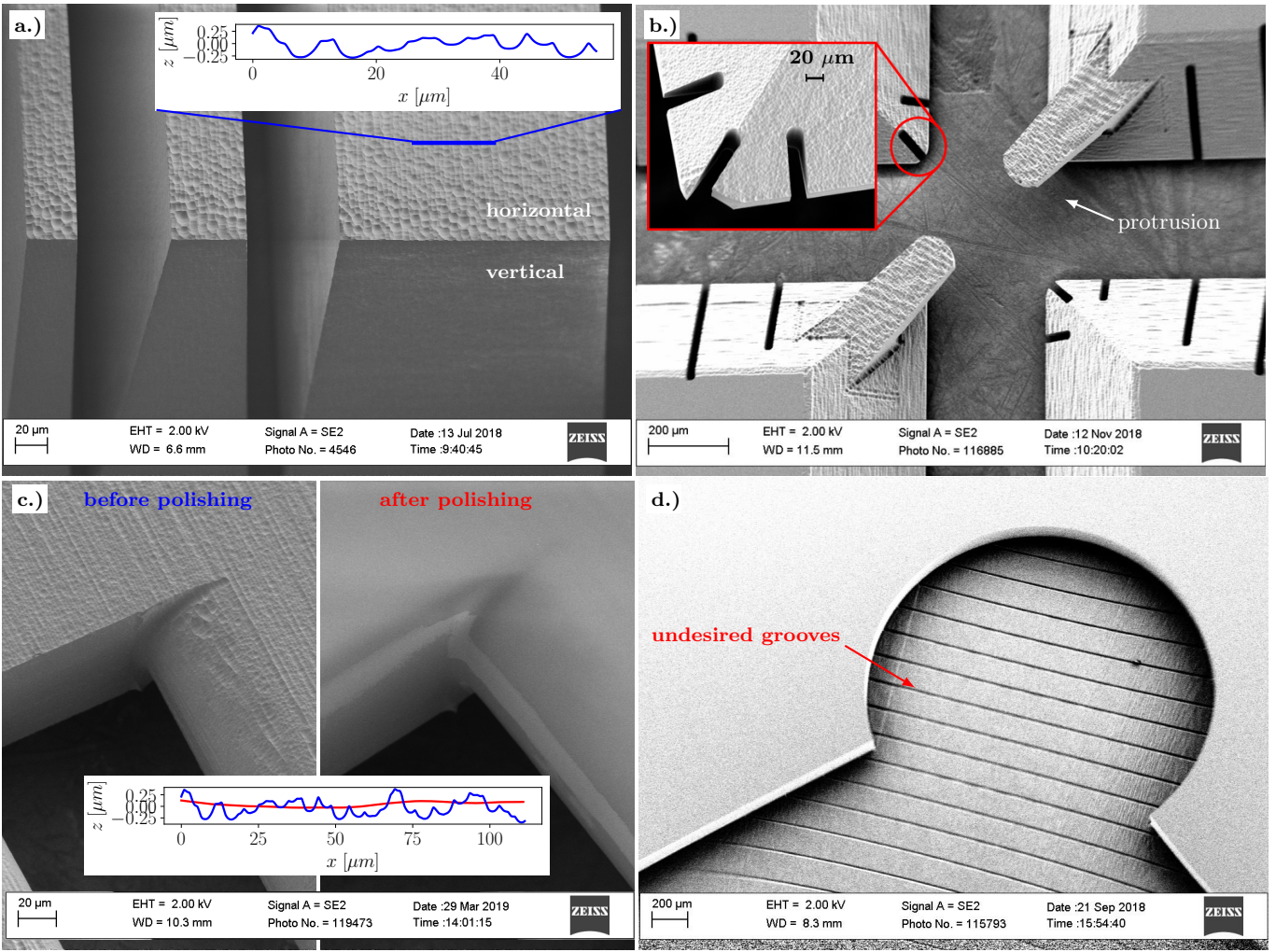
\twocolumngrid

The aspect ratio we tested for cuts and grooves was up to 1:50, which is sufficient for our applications, although the manufacturer quotes an achievable aspect ratio of up to 1:500 \cite{FEMTOprintWeb}. 
Our test samples have shown that cuts feature very regular and sharp edges (Figure \ref{fig:SEMImageFingerTestSample} \textbf{a.)}). From SEM images we estimate the radii of curvature of the edges to be less than 100 nm. We also observe that the femtosecond laser enhanced etching is capable of forming 3D shapes with sizes as small as 10 $\mu\mathrm{m}$ including freely protruding structures such as those shown in \ref{fig:SEMImageFingerTestSample} \textbf{b.)}. The laser-enhanced etching process is automized \cite{FEMTOprintWeb}, which in our experience results in short turnaround times on the order of a few weeks and thus provides the possibility for relatively fast prototyping.

\subsection{Limitations of the machining process}\label{Sec:LimitationsMachiningProcess}

While the versality of the laser-enhanced etching process is attractive, it is important to consider a number of factors which lead to imperfections. The first is that due to the shape of the LAZ and the nature of volume sampling the surface quality differs on horizontal and vertical surfaces. Panel \textbf{a.)} of Figure \ref{fig:SEMImageFingerTestSample} shows the top side of a sampled volume (surfaces perpendicular to the laser beam), for which the etching leads to a characteristic surface roughness on the order of hundreds of nanometers. Also shown are the sides where the surface roughness is only on the order of a few tens of nanometers. Surfaces with high roughness can be smoothened using an additional polishing step in which the femtosecond laser is re-used to melt the surface of the machined structure. Due to surface tension the surface becomes very smooth with micro roughness on the order 1 nm and a flatness that is still on the order of 10 - 100 nm. The result of polishing is shown in panel \textbf{c.)} of Figure \ref{fig:SEMImageFingerTestSample}. However, besides the desired smoothing of surfaces, structures such as edges and steps are also softened by this thermal polishing step.

A second challenge arises when large quantities of substrate material are removed. During intense laser irradiation  local heating can occur in the silica, which can lead to a slight bending or expansion of the sample during laser writing. As a consequence the overall process quality and precision is reduced. The removal of large quantities of material can also create stress in the sample. We have observed that this can result in small defects with sizes on the order of 10 to 20 $\mu$m. When working with 125 $\mu$m thick wafers we observed regular grooves on horizontal surfaces after etching, which did not occur with thicker wafers of similar material. Defects as well as grooves are especially likely to occur at positions where a large fraction of the surrounding material has been removed. Examples thereof from our junction trap are shown in Figure \ref{fig:SEMImageFingerTestSample} \textbf{b.)} and \textbf{d.)} and the inset of \ref{fig:SEMImageFingerTestSample} \textbf{b.)}. If these defects can not be tolerated, they limit the process yield since the occurrence is unpredictable. The removal of large quantities of material also requires a long laser irradiation step, which increases the process cost.

\section{Multi-layer trap design concept}\label{Main:MultiLayerDesignConcept}

We use the femtosecond laser-enhanced etching to develop ion trap wafers that offer simple and high precision wafer-to-wafer alignment. Our alignment strategy utilizes the precision machining to realize a three point support mechanism with the relevant alignment features directly machined into the individual wafers. These then self-align when placed on top of each other. In the following we describe this alignment scheme and establish general guidelines for the design of similar traps.

\begin{figure}[t]
    \includegraphics[width=0.486\textwidth]{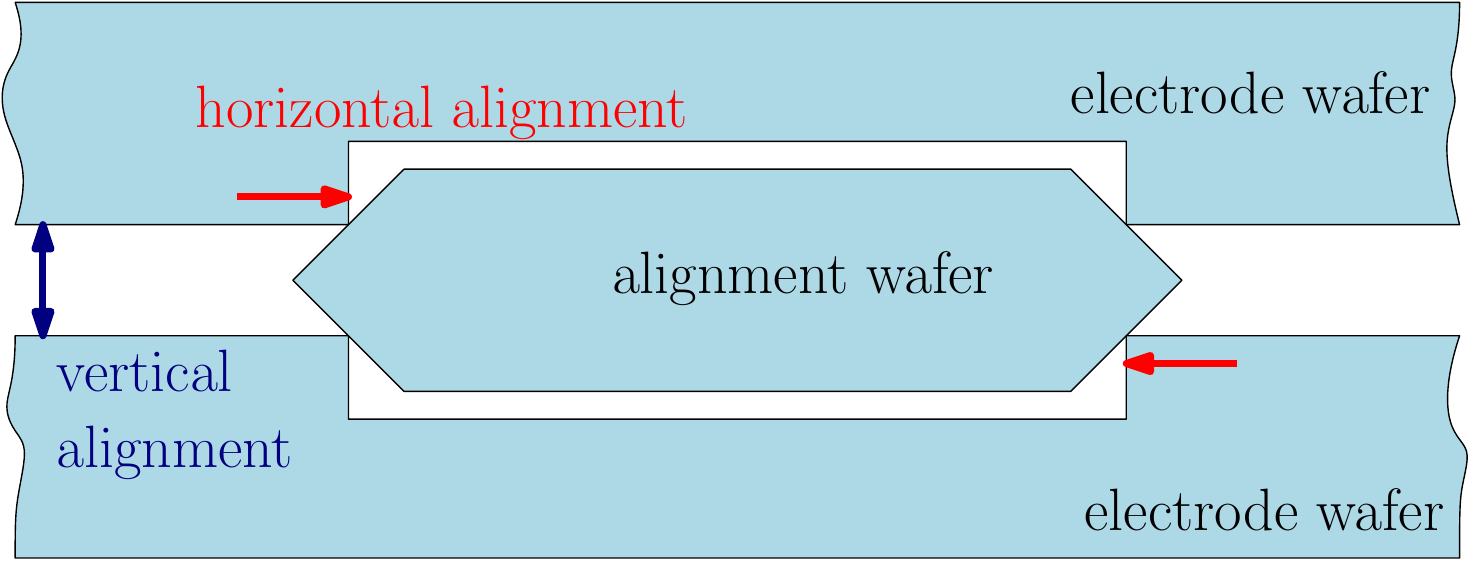}
    \put(-242,82){\crule[white]{.50cm}{.4cm}}
     \put(-240,85){\footnotesize{\textcolor{black}{\textbf{a.)}}}}\\
    \includegraphics[width=.486\textwidth,trim={6cm 5.cm 6.5cm 5.cm},clip]{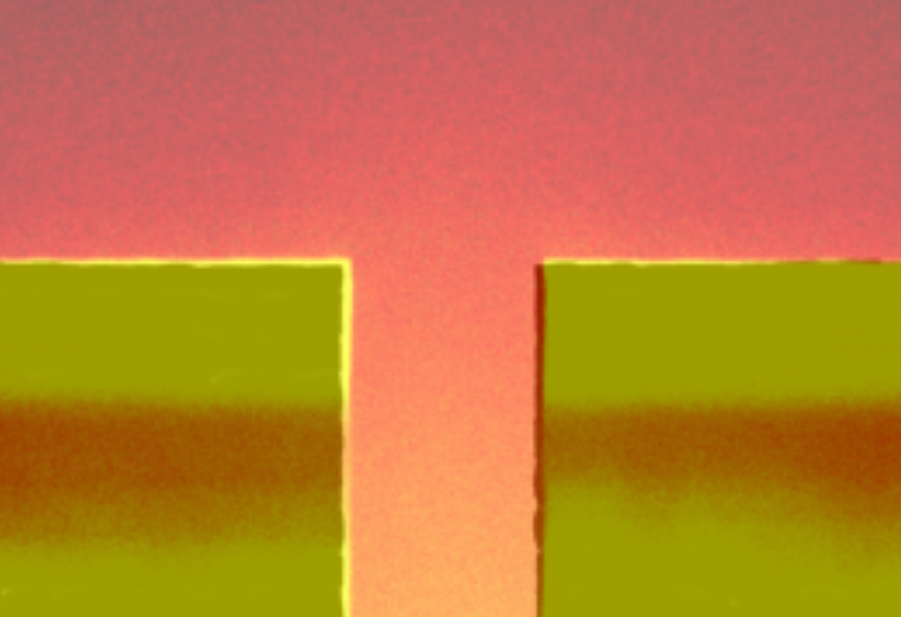}
    \put(-157,100){
    \begin{tikzpicture}
        \draw[black,thick](0,0) -- (0,0.75);
    \end{tikzpicture}}
    \put(-154,100){
    \begin{tikzpicture}
        \draw[black,thick](0,0) -- (0,0.75);
    \end{tikzpicture}}
    \put(-150,98){
    \begin{tikzpicture}
        \draw[black,thick](0,0) -- (0.75,0);
    \end{tikzpicture}}
    \put(-150,96){
    \begin{tikzpicture}
        \draw[black,thick](0,0) -- (0.75,0);
    \end{tikzpicture}}
     \put(-140,108){\textcolor{black}{$\lesssim\mathbf{2\ \mathrm{\mu m}}$ mismatch}}
     \put(-247,157){\crule[white]{.50cm}{.4cm}}
     \put(-245,160){\footnotesize{\textcolor{black}{\textbf{b.)}}}}
     \put(-38,15){
       \begin{tikzpicture}
         \draw[black,line width=0.6mm](0,0) -- (.95,0);
        \end{tikzpicture}}
     \put(-38,11){
      \begin{tikzpicture}
        \draw[black,line width=0.6mm](0,0) -- (0,0.3);
      \end{tikzpicture}}
     \put(-11,11){
      \begin{tikzpicture}
        \draw[black,line width=0.6mm](0,0) -- (0,0.3);
      \end{tikzpicture}}
     \put(-35,23){\textcolor{black}{$\mathbf{10\ \mathrm{\mu m}}$}}\\
    \includegraphics[width=.486\textwidth,trim={1.5cm 7cm 0cm 7cm},clip]{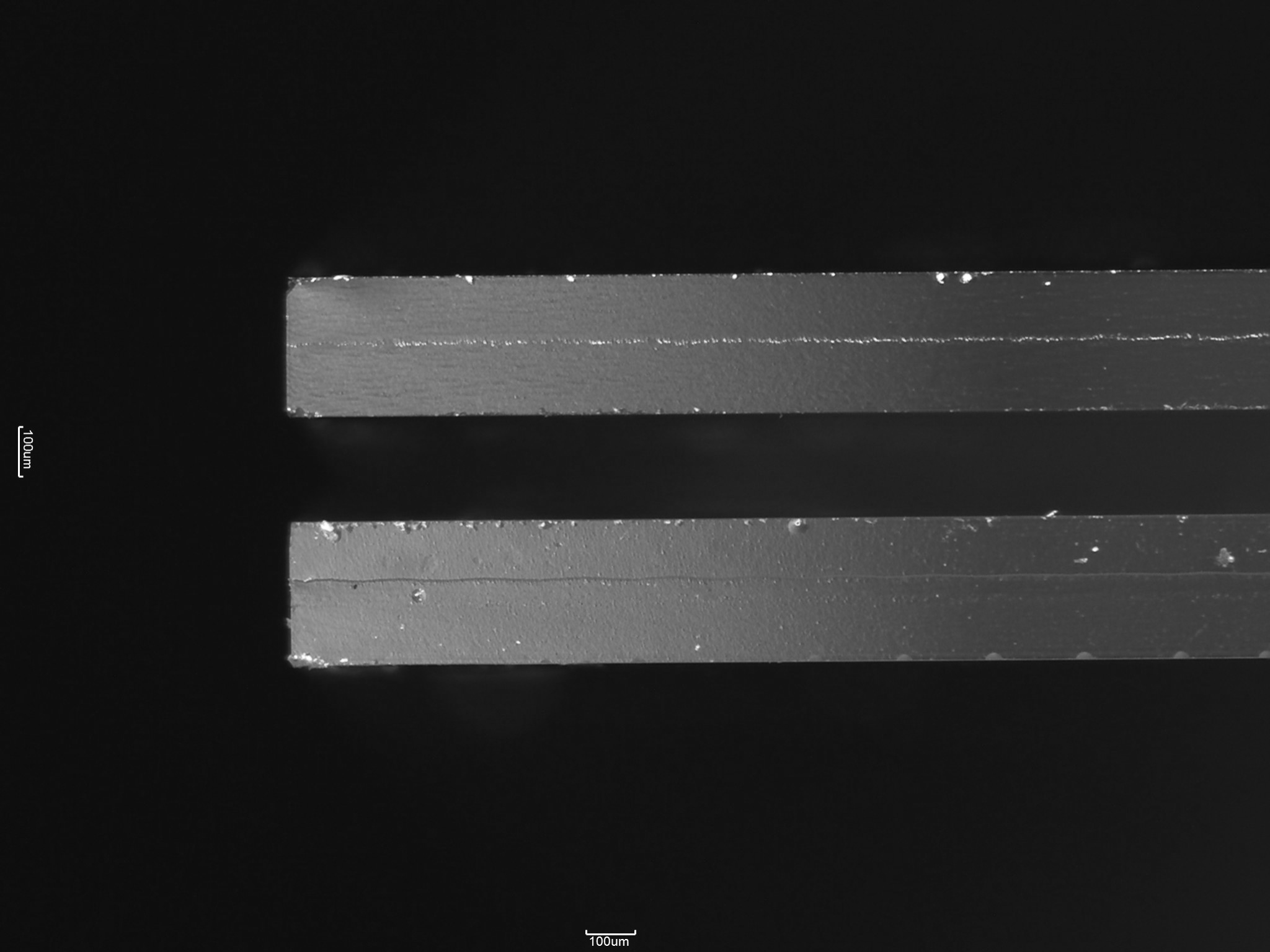}
     \put(-70,90){\textcolor{white}{top wafer}}
     \put(-70,40){\textcolor{white}{bottom wafer}}
     \put(-180,65){\textcolor{white}{wafers parallel to less than 0.05 deg}}
     \put(-247,118){\crule[white]{.50cm}{.4cm}}
     \put(-245,121){\footnotesize{\textcolor{black}{\textbf{c.)}}}}
          \put(-243,10){
       \begin{tikzpicture}
         \draw[white,line width=0.5mm](0,0) -- (.70,0);
        \end{tikzpicture}}
     \put(-243,6){
      \begin{tikzpicture}
        \draw[white,line width=0.5mm](0,0) -- (0,0.3);
      \end{tikzpicture}}
     \put(-223,6){
      \begin{tikzpicture}
        \draw[white,line width=0.5mm](0,0) -- (0,0.3);
      \end{tikzpicture}}
     \put(-240,18){\textcolor{white}{$\mathbf{200\ \mathrm{\mu m}}$}}\\
    \caption{\textbf{a.)} Cross-section through a stack of three wafers with integrated alignment. \textbf{b.)} Measurement of the horizontal alignment precision of two electrode wafers in a wafer stack with integrated alignment. The stack is imaged from the top, the mismatch is found to be less than $2\ \mu\mathrm{m}$, limited by the depth of view and precision of mechanical parts of the microscope. \textbf{c.)} Stacked trap wafers imaged from the side with a microscope. The distance between the two wafers is within 2 micron on all 4 corners of the stack.}
    \label{fig:AlignmentCrossSection}
\end{figure}
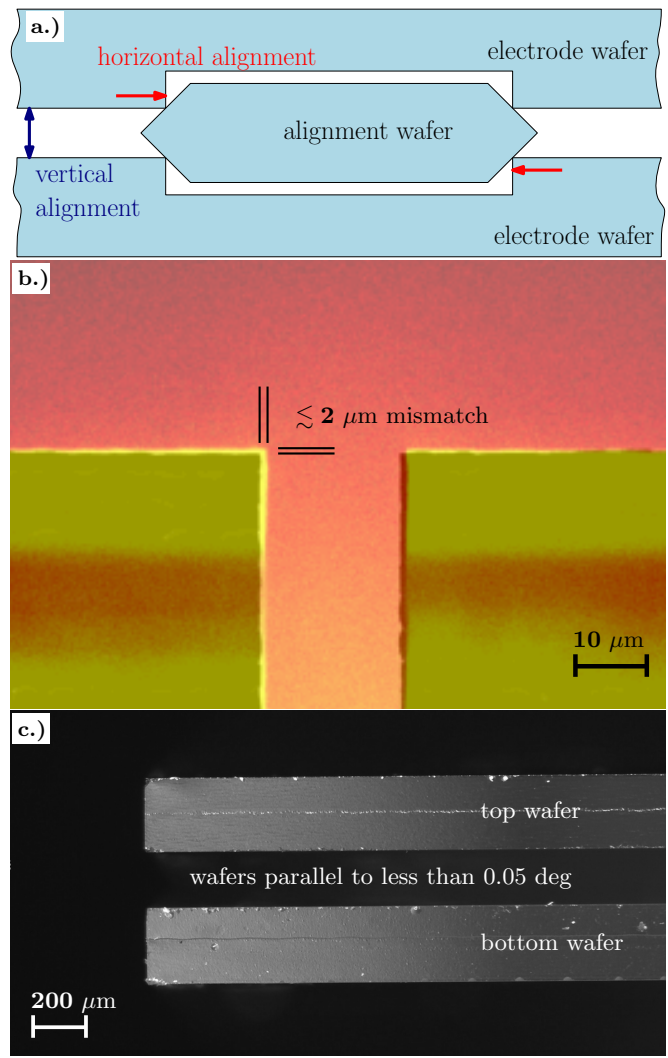

\subsection{Alignment strategy}\label{Sec:AlignmentStrategy}

Our alignment scheme is based on a three point support mechanism, which mechanically defines the relative orientation of neighbouring wafers. In the two traps described below, we have used a stack of three wafers, with the middle wafer defining the three support points for the other two. Any of these wafers might be used for electrodes. In our assembly process the wafers are placed together with a small amount of clamping force, and then glued in place. 

The design of the alignment points is illustrated in the cross-section sketch in \ref{fig:AlignmentCrossSection} \textbf{a.)}. We found that the alignment points on the middle wafer work well if they are half circles with $45^{\circ}$ chamfer on the top and bottom side each half way through the wafer. A first approach with fully integrated alignment features failed due to poor precision and low surface quality (see appendix \ref{App:WaferAlign}).

The respective counterpart for the alignment on the outer wafers is an indentation (see Figure \ref{fig:ElectrodeDesignCrossSection} \textbf{a.)}). The walls of the indentation are vertical and make a right angle with the top surface of the wafer.
The size (and shape) of the inset is designed such that electrode and alignment wafers only touch each other at the alignment points. With the high precision attained by the machining, we achieve stable and precise alignment of the three wafers.

Figure \ref{fig:AlignmentCrossSection} \textbf{b.)} shows a measurement of the relative alignment of two wafers which are separated by an intermediate spacer in the manner described. The measurement is carried out using an optical microscope equipped with a CCD-camera. First the microscope is focused on an electrode of the bottom wafer of the wafer-stack and a picture is taken. Next, the microscope is focused on the same electrode on the top wafer and another picture is taken. Then the two pictures are overlapped pixel by pixel (both with increased contrast and one with inverted colors to obtain bright areas appearing dark). From the known pixel size and microscope magnification we can estimate the misalignment between the wafers. The measurement is limited by the depth of view and precision of mechanical parts of the microscope that could shift the images during refocusing. Nevertheless it allows us to constrain the relative alignment imprecision to $< 2$~$\mu$m. The stated alignment precision is for wafers of dimension of $20\times20\ \mathrm{mm}$. Similar measurement results were obtained at different points distributed over the full trap. 

A measurement of the vertical alignment is shown in panel \textbf{c.)} of Figure \ref{fig:AlignmentCrossSection}. We have observed that vertically the angle between the wafers is less than 0.05 deg while the separation relies on the thickness tolerance of the substrates. This typically lies between $\pm\ 5\ \mathrm{\mu m}$ and $\pm\ 20\ \mathrm{\mu m}$. By measuring the thickness of the alignment wafer prior to fabrication, the alignment feature on the electrode wafers could be adjusted to compensate for thickness deviation and thus the vertical alignment of the wafers could be improved if desired.  Within individual wafers, we have measured a homogeneous flatness over the full area of $20\times 20$ mm within uncertainty ($\pm 5\ \mathrm{\mu m}$) of the instrument. Even when wafer flatness varies by a few microns over the whole area, placing the alignment points far from each other ensures only a small variance in the angle between electrodes on each wafer. 

\subsection{Wafer and electrode design}\label{Sec:WaferElectrodeDesign}

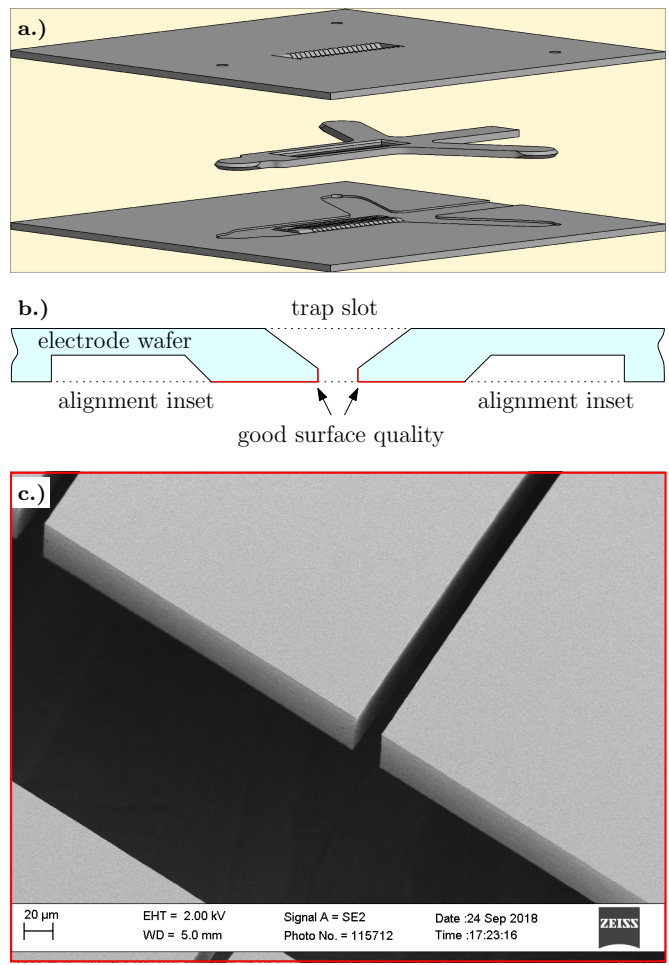
\begin{figure}[t]
    \begin{center}
    \includegraphics[width=0.486\textwidth]{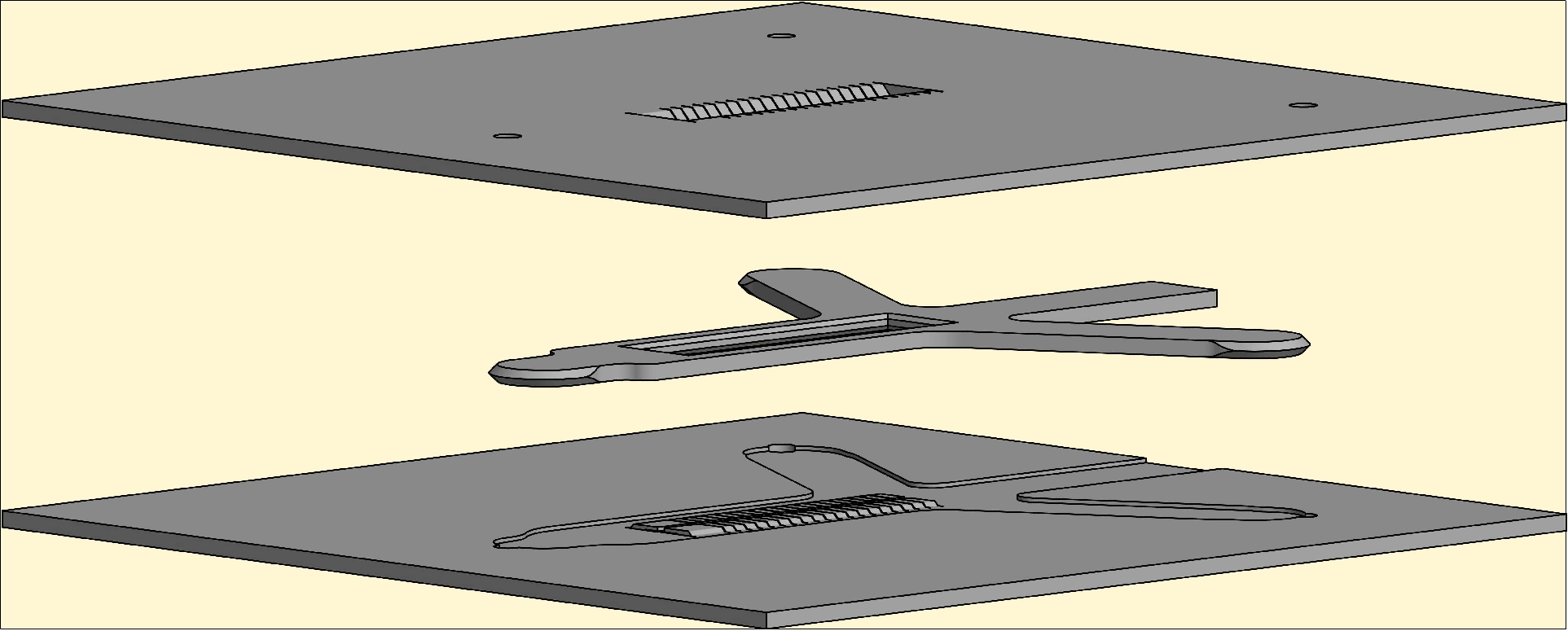}
    \put(-245,90){\footnotesize{\textcolor{black}{\textbf{a.)}}}}\\
    \includegraphics[width=0.486\textwidth]{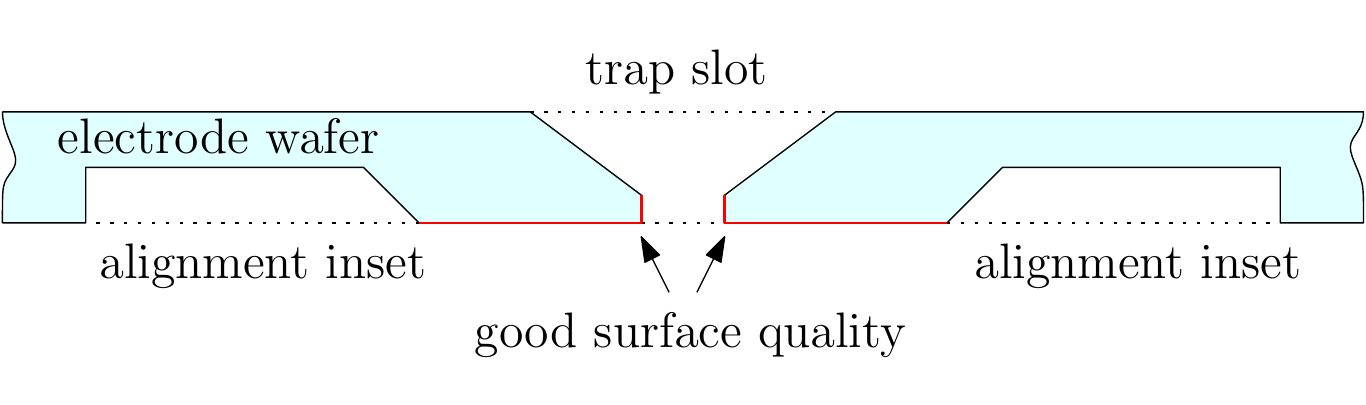}
    \put(-247,55){\crule[white]{.50cm}{.4cm}}
    \put(-245,58){\footnotesize{\textcolor{black}{\textbf{b.)}}}}\\    \includegraphics[width=.486\textwidth]{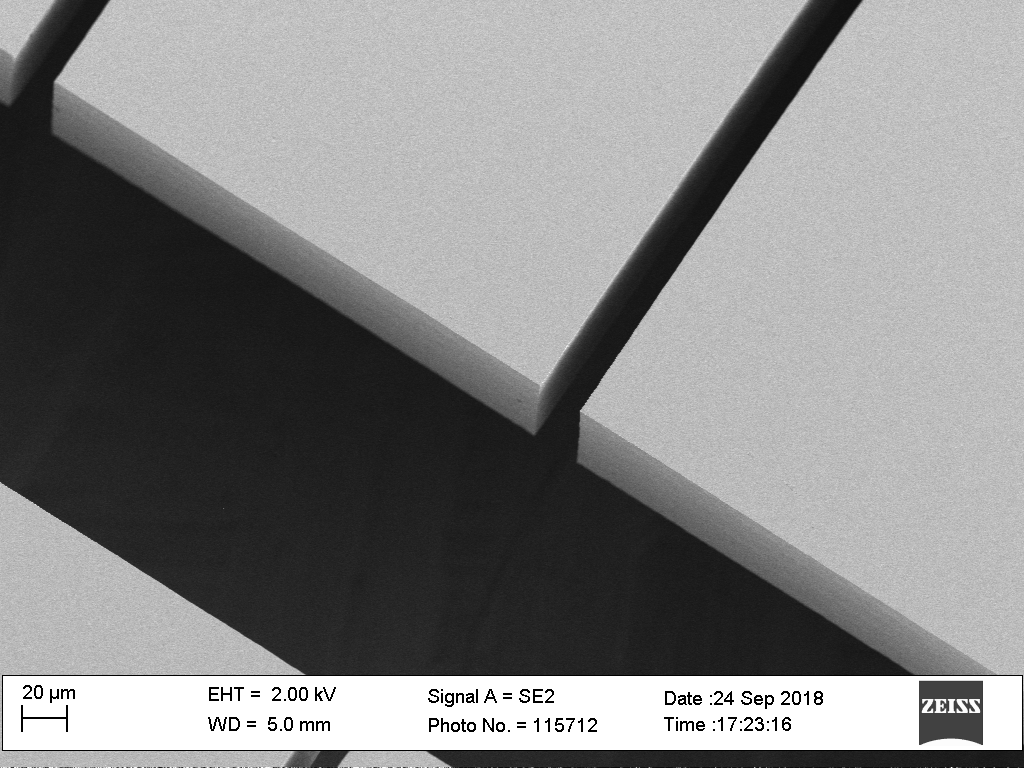}
     \put(-251,0){
\begin{tikzpicture}
    \draw[red,thick] (9.7,7.5) rectangle (1,1);
\end{tikzpicture}}
    \put(-247,172){\crule[white]{.50cm}{.4cm}}
    \put(-245,175){\footnotesize{\textcolor{black}{\textbf{c.)}}}}
    \caption{\textbf{a.)} Illustration of the full trap design. Shape of the middle wafer is optimized for low surface area. \textbf{b.)} Cross-section through the electrode configuration in the trap center. The dashed  lines indicate where material was removed during laser-machining. The red lines mark unmachined surfaces of high quality. These surfaces are close to the trap center. \textbf{c.)} SEM image of the electrodes of a sample trap. The surfaces shown are the ones indicated by red color in the sketch in panel \textbf{b.)}. The surface roughness is in the tens of nm range or below which should ensure low anomalous heating rates.}
    \label{fig:ElectrodeDesignCrossSection}
    \end{center}
\end{figure}

One possible wafer design is illustrated in Figure \ref{fig:ElectrodeDesignCrossSection} \textbf{a.)} and \textbf{b.)}. For this design the trap is positioned in the center of the wafers between the three alignment points. The middle wafer features a slot in the center where the trap is located. Furthermore, the shape of the middle wafer allows the electrodes defined on the top and bottom wafer to extend towards the trap center (see Figure \ref{fig:ElectrodeDesignCrossSection} \textbf{b.)} and \ref{fig:alignment_chiara} \textbf{a.)}). In this way, the electrodes are defined on a non-machined surface and vertical cuts through the wafer define the trap area and segmentation of the electrodes. An example of machined electrodes is shown in Figure \ref{fig:ElectrodeDesignCrossSection} \textbf{c.)}. The silica electrodes show sharp regular edges and very little surface roughness on the vertical faces of the segments, which is beneficial for building a small trap with low ion-electrode distance.\\
A judicious choice of initial wafer thickness is advantageous, since it prevents adverse effects associated with the removal of a large amount of material.

\section{Specific trap designs}\label{Sec:TrapDesign}

\begin{figure}[H]
\centering
\begin{subfigure}{\textwidth}
\begin{subfigure}{0.5\textwidth}
  \includegraphics[width=.49\textwidth,trim={0cm 0 0cm 0},clip]{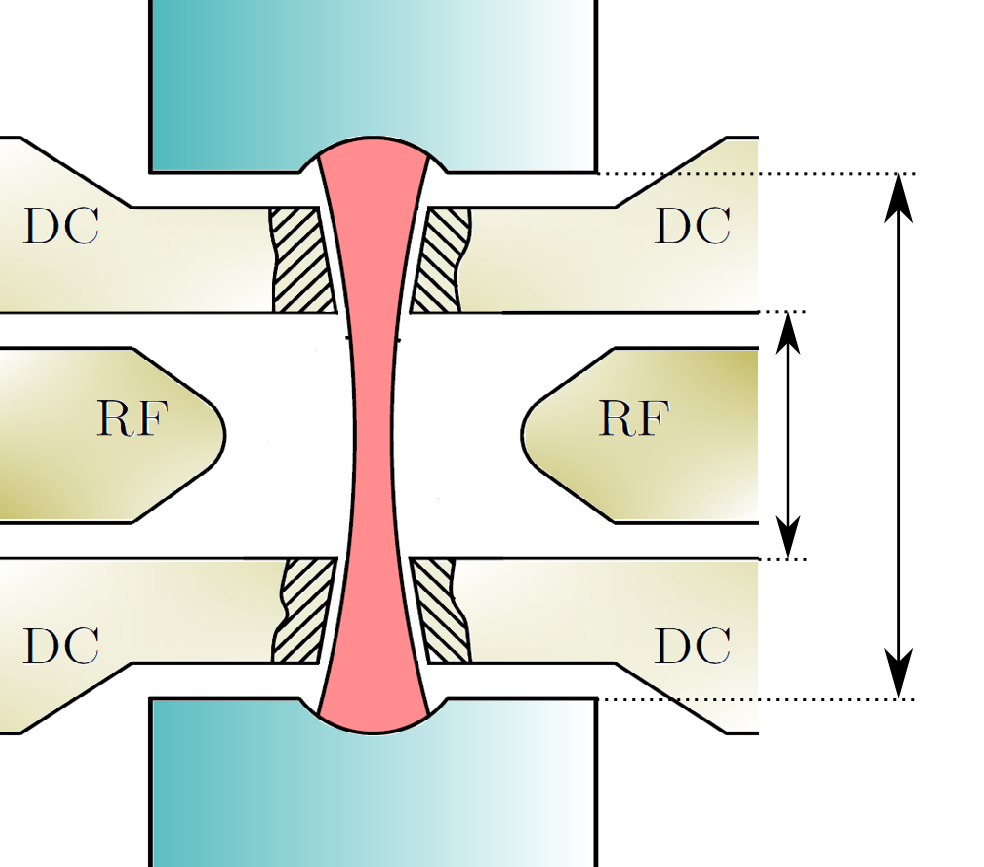}
  \includegraphics[width=.50\textwidth,trim={7.5cm 4.5cm 2.5cm 0cm},clip]{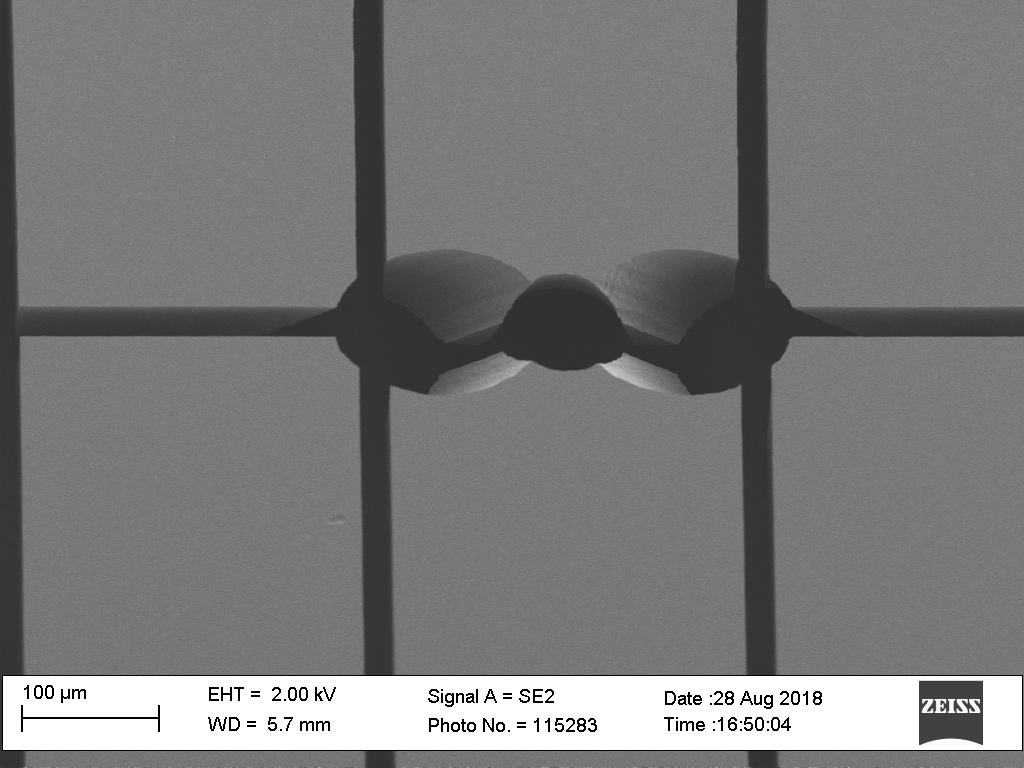}
  \put(-125,8){
     \begin{tikzpicture}
         \draw[white,line width=0.5mm](0,0) -- (.75,0);
   \end{tikzpicture}}
  \put(-125,6){
     \begin{tikzpicture}
         \draw[white,line width=0.5mm](0,0) -- (0,0.15);
   \end{tikzpicture}}
  \put(-104,6){
     \begin{tikzpicture}
         \draw[white,line width=0.5mm](0,0) -- (0,0.15);
   \end{tikzpicture}}
  \put(-120,14){\footnotesize{\textcolor{white}{100 $\mu$m}}}
  \put(-255,105){\footnotesize{\textcolor{black}{\textbf{a.)}}}}
  \put(-233,100){\footnotesize{\textcolor{black}{\textbf{fiber cavity}}}}
  \put(-150,40){\scriptsize{\rotatebox{90}{\textcolor{black}{\textbf{$170\ \mathrm{\mu m}$}}}}}
    \put(-138,40){\scriptsize{\rotatebox{90}{\textcolor{black}{\textbf{$350\ \mathrm{\mu m}$}}}}}
  \put(-125,97){\crule[white]{.5cm}{.4cm}}
  \put(-123,100){\footnotesize{\textcolor{black}{\textbf{b.)}}}}\\
 \includegraphics[width=\textwidth]{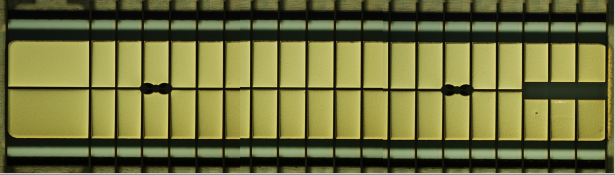}
  \put(-36,4){\crule[white]{1.25cm}{.25cm}}
  \put(-86,4){\crule[white]{1.55cm}{.25cm}}
  \put(-201,4){\crule[white]{1.1cm}{.25cm}}
  \put(-252,57){\crule[white]{.5cm}{.4cm}}
  \put(-35,5){\footnotesize{\textcolor{black}{\textbf{loading}}}}
  \put(-85,5){\footnotesize{\textcolor{black}{\textbf{detection}}}}
  \put(-200,5){\footnotesize{\textcolor{black}{\textbf{cavity}}}}
  \put(-250,60){\footnotesize{\textcolor{black}{\textbf{d.)}}}}
\end{subfigure}
\begin{subfigure}{0.5\textwidth}
  \includegraphics[width=\textwidth]{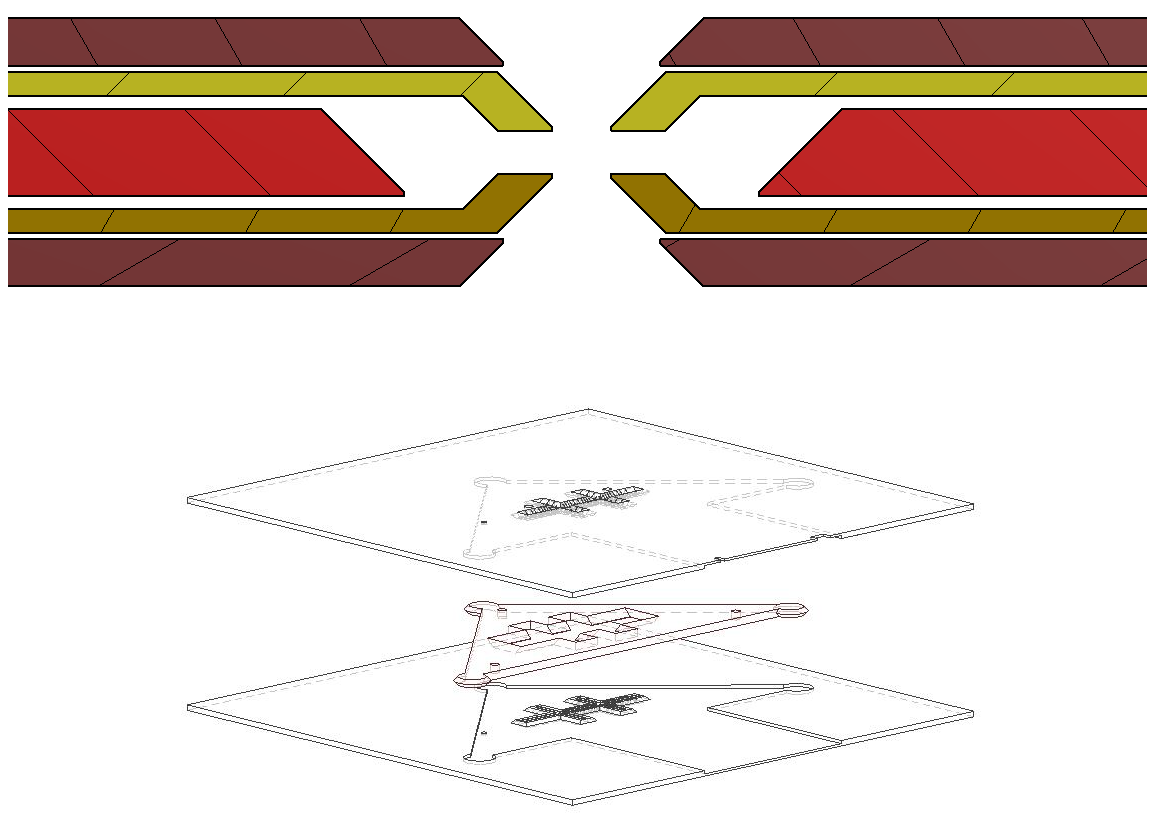}
  \put(-251,167){\crule[white]{.5cm}{.4cm}}
  \put(-249,170){\footnotesize{\textcolor{black}{\textbf{c.)}}}}
  \put(-249,105){\footnotesize{\textcolor{black}{\textbf{e.)}}}}
\end{subfigure}
\end{subfigure}
\caption{\textbf{a.)} Cross-section illustration of the cavity zone, cavity length $350\ \mathrm{\mu m}$. \textbf{b.)} SEM image of the holes for laser access in one of the DC-wafers. \textbf{d.)} Shows a capture of one DC-wafer (constructed from four microscope images) over the full length of the trap with cavity zone on the left, detection zone and loading zone on the right. Segmented DC-electrode for shuttling the ions along the trap axis, width of one segment: $250\ \mu\mathrm{m}$. \textbf{c.)} Side view of the full wafer stack of the double junction trap, main trapping wafers are shown in yellow, compensation wafers in dark red and middle wafer in bright red. \textbf{e.)} stack of the main 3 wafers, the outer wafers carry RF and DC electrodes, the inner wafer serves as an alignment piece and can be used to integrate optical fibres.}
\label{fig:ImageCavityIntTrap}
\label{fig:alignment_chiara}
\end{figure}
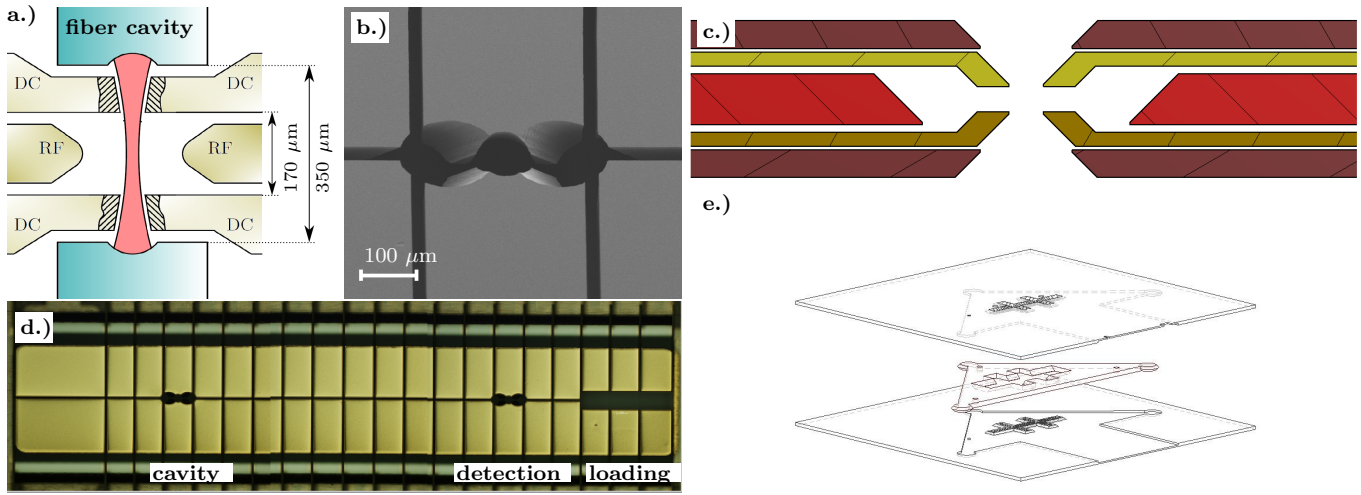

In this section we present two trap designs which aim to investigate aspects of scaling trapped-ion control for 
quantum information processing. The first trap design is a linear segmented trap which was designed to be compatible with a short-length optical cavity for enhanced ion-photon coupling \cite{Takahashi2018}. One primary concern in designing a trap integrated with an optical cavity is that the dielectric coatings of the mirrors can accumulate stray charges which disturb the ions \cite{Harlander2010}. Thus, shielding of stray fields is highly desirable. The machining precision provided by the laser-assisted etching described above allows to build a small multi-layer trap, in which electrodes can be placed in-between the cavity mirrors while keeping the cavity length short - in this case we aim for a cavity of length 300~$\mu$m. A cross-section illustration of the proposed cavity region is shown in Figure \ref{fig:ImageCavityIntTrap} \textbf{a.)}. The trap design is based on three layers; two layers with DC-electrodes and one with RF-electrodes which is sandwiched between the other two, also serving as alignment wafer.

In our trap prototype we aim for an ion to nearest electrode distance of around $90\ \mu\mathrm{m}$. Access for laser beams and the cavity mode is provided through holes with $100\ \mu\mathrm{m}$ diameter in the electrodes, which otherwise block all optical access. This provides shielding against stray fields originating outside the electrode structure. A close up view of these holes is shown in Figure \ref{fig:ImageCavityIntTrap} \textbf{b.)}. The trap features two regions suited for optical cavities or optical fibers for light collection and delivery. It also includes an open ``loading'' region so that an atomic beam can be introduced for loading ions. A view of the DC electrodes over the full length of the trap is shown in Figure \ref{fig:ImageCavityIntTrap} \textbf{d.)}. The ions will be shuttled along the trap axis using the segmented electrodes.

The second trap we have designed and fabricated is a 3-dimensional segmented double junction trap. The trap consists of a total stack of 5 wafers, as shown in Figure \ref{fig:alignment_chiara}\textbf{c.)}: a middle wafer which serves as a spacer and alignment piece, two outer wafers which carry the main DC and RF trapping electrodes, and two additional wafers carrying electrodes for stray field compensation. The three central wafers, which are the inner ones in the full stack, are shown in Figure \ref{fig:alignment_chiara} \textbf{e.)}. All wafers are self aligned in a similar fashion as described in section III A, and subsequently glued to each other.
This trap has 144 electrodes, and includes many independently addressable experimental zones as well as splitting, junction and storage zones. The two X junctions are key ingredients for ion transport into 2-dimensions, which is critical for scaling. Several features rely on the ability to form 3-dimensional structures in fused silica. One such feature is the RF electrode shape at the junction. Other features which we have implemented are electrodes angled at 45 degrees and vias for electrode connectivity. In initial designs we also included deep grooves which could host integrated optical fibres. We will briefly describe each in the following. 

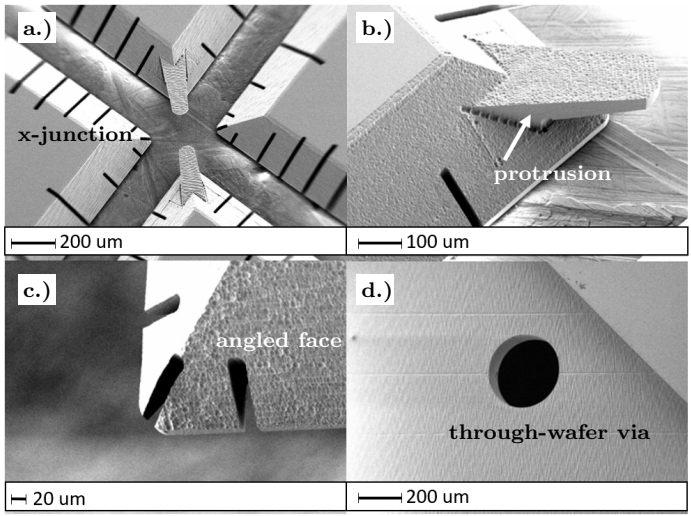
\begin{figure}[t!]
\centering
\includegraphics[scale=0.35]{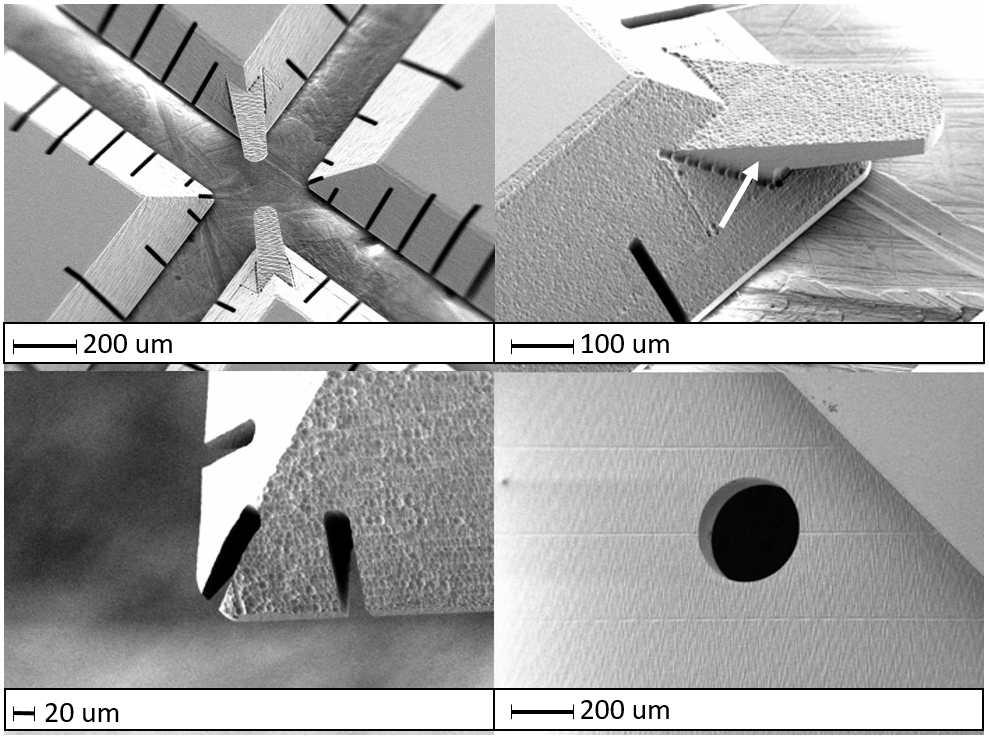}
    \put(-255,180){\crule[white]{.50cm}{.4cm}}
    \put(-253,183){\footnotesize{\textcolor{black}{\textbf{a.)}}}}
    \put(-255,143){\footnotesize{\textcolor{black}{\textbf{x-junction}}}}
    \put(-127,180){\crule[white]{.50cm}{.4cm}}
    \put(-125,183){\footnotesize{\textcolor{black}{\textbf{b.)}}}}
    \put(-75,128){\footnotesize{\textcolor{white}{\textbf{protrusion}}}}
    \put(-255,81){\crule[white]{.50cm}{.4cm}}
    \put(-253,84){\footnotesize{\textcolor{black}{\textbf{c.)}}}}
    \put(-180,65){\footnotesize{\textcolor{white}{\textbf{angled face}}}}
    \put(-127,81){\crule[white]{.50cm}{.4cm}}
    \put(-125,84){\footnotesize{\textcolor{black}{\textbf{d.)}}}}
    \put(-92,30){\footnotesize{\textcolor{black}{\textbf{through-wafer via}}}}
\caption{Features of the junction trap which are made possible by the enhanced laser writing technology: \textbf{a.)} and \textbf{b.)} protruding 3D structures serving as partial RF bridges, \textbf{c.)} surfaces angled at 45 degrees for enhanced optical access and \textbf{d.)} through-wafer via for electrical connectivity. }
\label{fig:chia_feat}
\end{figure}

At the corners of the electrodes at the junction, two opposing protrusions  (Figure \ref{fig:chia_feat} \textbf{a.)} and \textbf{b.)}) are used to break the symmetry of electric fields, producing 3-dimensional confinement of an ion at the center of the junction, which is not present for a 4-fold symmetric design \cite{BlakestadNIST2010}. A previous junction trap was built at NIST, but featured full bridges which crossed from one side of the junction to the other \cite{BlakestadNIST2010}. Both the protrusions and bridges give rise to pseudopotential energy barriers which the ion must traverse as it enters the junction. However these are much larger for a full bridge than for the protrusions.
\onecolumngrid

\twocolumngrid
Gradients of the pseudopotential at the sides of the barriers can introduce undesirable heating mechanisms. With the flexibility available from the laser-enhanced etching, we are able to chose the length of the protrusions - the final design was chosen as a compromise between confinement at the center and the magnitude of the energy barriers.  The resulting protrusions are 100 $\mu$m wide and 50 $\mu$m thick.

The front surface of all electrodes is angled at 45 degrees to provide high optical access for laser beams and a large solid angle for light collection. Since these faces are machined they exhibit a roughness of hundreds of nm, as visible in Figure \ref{fig:chia_feat} \textbf{c.)}. This roughness is not on the side of the trap that faces the ion, therefore any imperfections on this face should be shielded at some level from the ion. The RF electrode is defined on the inner (ion-facing) side of the wafer, and then is connected through a via (Figure \ref{fig:chia_feat} \textbf{d.)}) to the top wafer surface where it is wire bonded. The via consists of a 0.3 mm diameter hole with vertical faces. The electrical connection is made by evaporating gold on it at an angle. We have investigated the resistance of the electrode when evaporated with a thin layer of gold (700 nm) and have detected an increase of the resistance through the via on the order of 1 Ohm. Typically, evaporated tracks in the range of 10 mm in length (as are often used in electrode connections)  exhibit a resistance between 1 and 5 Ohms. A subsequent electroplating step should allow to further reduce this resistance.  Different via diameters are also possible. Tests on a 300 $\mu$m thick wafer have shown that holes can be machined reliably down to a 50 $\mu$m diameter. As we can fully coat gaps with width on the order of 20 $\mu$m and similar vertical depth, we think that it should be possible to fully coat 50 $\mu$m vias.
\begin{figure}[t!]
\centering
\includegraphics[scale=0.45]{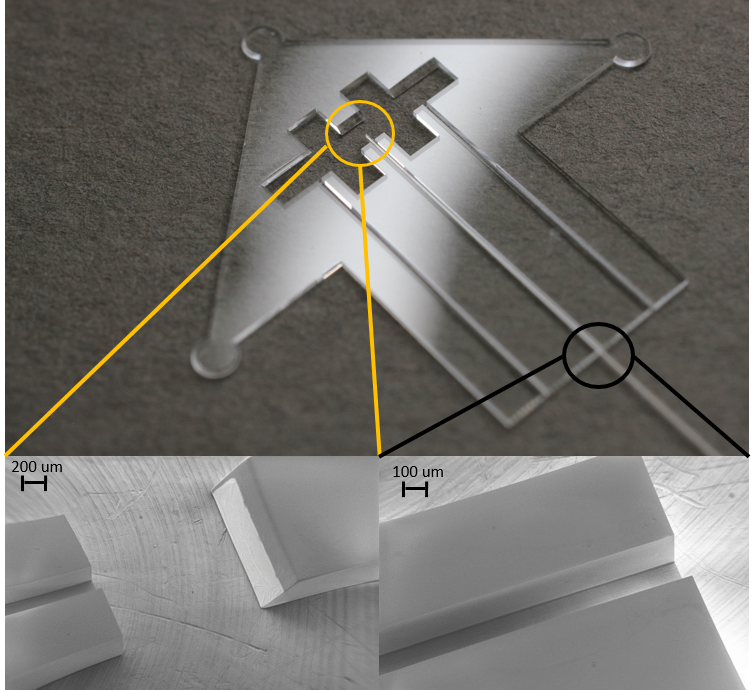}
    \put(-248,2){\crule[white]{.50cm}{.4cm}}
    \put(-246,4){\footnotesize{\textcolor{black}{\textbf{a.)}}}}
    \put(-121,2){\crule[white]{.50cm}{.4cm}}
    \put(-119,4){\footnotesize{\textcolor{black}{\textbf{b.)}}}}
    \put(-230,21){\footnotesize{\textcolor{white}{\textbf{groove}}}}
    \put(-210,61){\footnotesize{\textcolor{black}{\textbf{polished mirror}}}}
      \put(-195,46){
  \begin{tikzpicture}[thick]
    \draw [black,   -latex      ] (0,0.3) -- (0.7,0.0) node [right] {};
  \end{tikzpicture}
  }
    \put(-80,21){\footnotesize{\textcolor{white}{\textbf{squared groove}}}}
\caption{Integration of lensed optical fibres within the middle wafers. Close to the trap center, a facet angled at 45 degrees is polished and metal-coated to create a mirror to direct the light out of the trap, as shown in inset \textbf{a.)}. Squared grooves to guide the fibers are machined in the middle wafer, as shown in inset \textbf{b.)}.}
\label{fig:optical_int}
\end{figure}
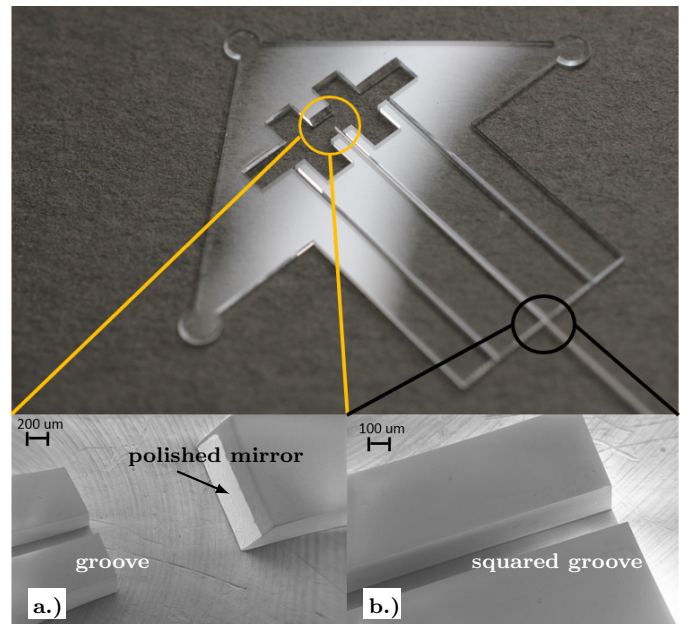

A further set of tests were performed to assess the viability of incorporating optical fibres for laser beam delivery. For this purpose, following the approach of \cite{Ott2016}, we fabricated squared grooves into the middle wafer which are shown in inset \textbf{b.)} of Figure \ref{fig:optical_int}. We found that these could be machined with horizontal widths which are uncertain to $\pm$ 5 $\mu$m. Our original aim in these tests was to insert photonic crystal optical fibres (PCF) in these low tolerance grooves, ensuring good vertical and horizontal alignment of the beams. PC fibres were stripped of their coating layer and their glass cladding, which has a 230 $\pm$ 5 $\mu$m diameter, and were inserted and glued in 237 $\mu$m wide grooves. For laser waist diameters between 30 and 40 $\mu$m, this technique provides a relatively easy and robust alignment, with maximal theoretical displacements of 6.5 $\mu$m between the ion's location and the center of the beam waist. The PC fibres used had lenses melted onto the tip \cite{WTTWeb}, with a working distance in the order of 500 $\mu$m at 729 nm. To allow light to exit the structure, we envisioned using polished mirror surfaces on the opposite side as the fibre grooves, directing incident light out of the trap structure, as shown in inset \textbf{a.)} of Figure \ref{fig:optical_int}. We have used an SEM to image the mirror surfaces machined using laser-enhanced etching, which, when gold coated, seem suitable for our purpose. Their micro-roughness is in the order of a nanometer, as shown in Figure \ref{fig:SEMImageFingerTestSample} \textbf{c.)}.

\begin{figure}[t]
\centering
\includegraphics[scale=0.5]{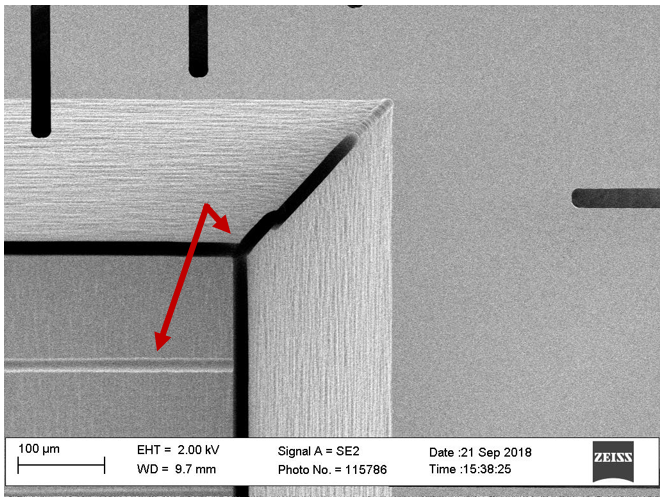}
    \put(-225,115){\textcolor{dred}{\textbf{undesired grooves}}}
\caption{Machining defects can appear in areas where a lot of material is removed, and where faces at different angles meet. Cuts up to 10 $\mu$m in depth can also be present, as indicated by the red arrows. These can be mitigated by introducing a small radius of curvature and by applying a polishing step at the end of the process.}
\label{fig:undesired_groove}
\end{figure}

The complex geometries present in the double junction trap produce unexpected stress in the silica substrate. As a consequence we have noticed an increased number of defects, particularly at the junction, where lots of material has been exposed to the laser. We have also observed stray machining cuts on several samples at the interface between different zones of the wafer, such as along the 45 degree angled faces, as shown in Figure \ref{fig:undesired_groove}, and on planar surfaces where material had been removed. These undesired grooves, which can reach 15 $\mu$m in depth, have been avoided by implementing a small radius of curvature at all intersecting surfaces and by applying a final polishing step following the machining. We conclude that it is important to consider these effects in the design stage.

\section{Fabrication process}\label{Sec:Fabrication}

\begin{figure}[H]
\centering
\begin{subfigure}{\textwidth}
  \includegraphics[width=.496\textwidth]{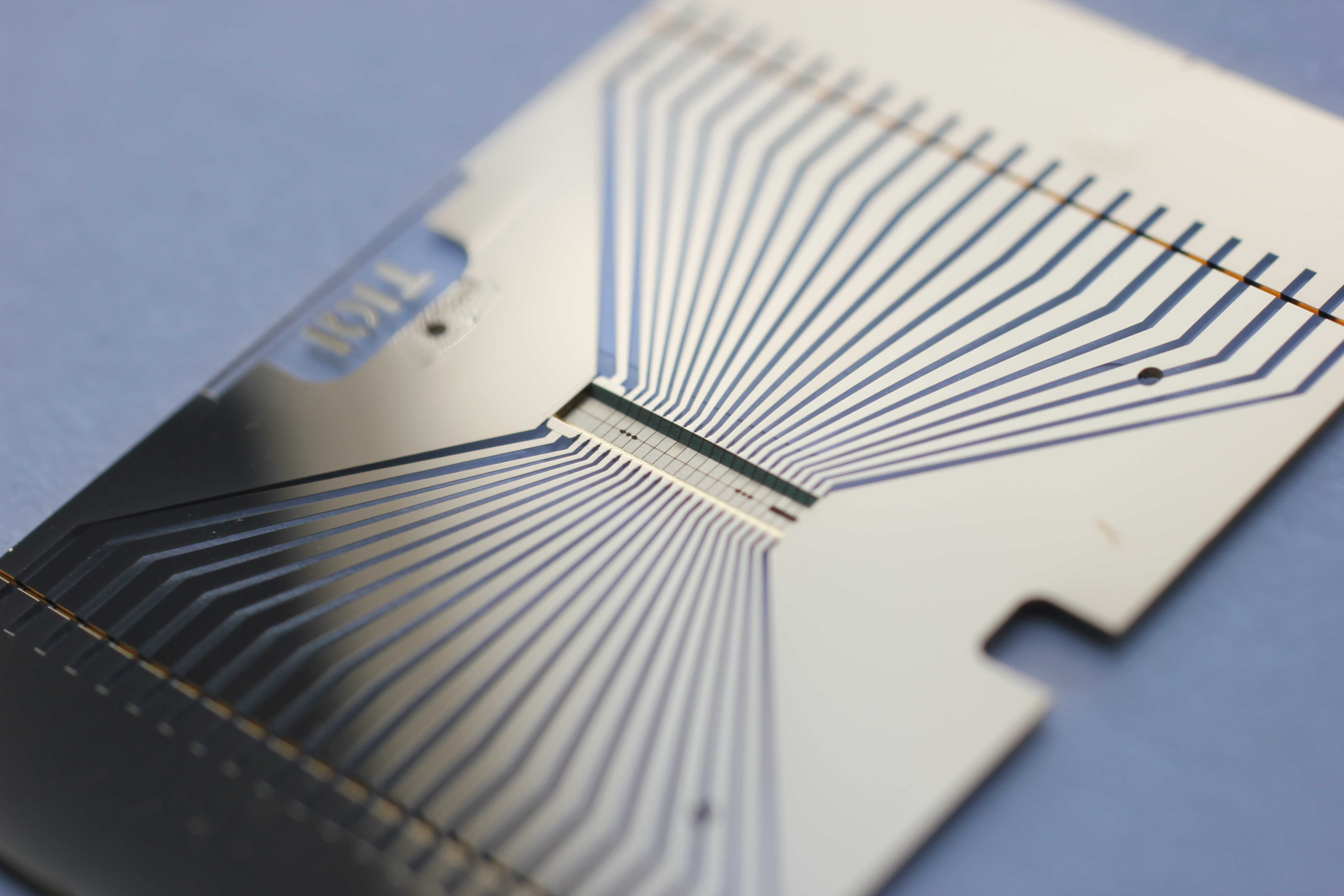}
  \includegraphics[width=.496\textwidth]{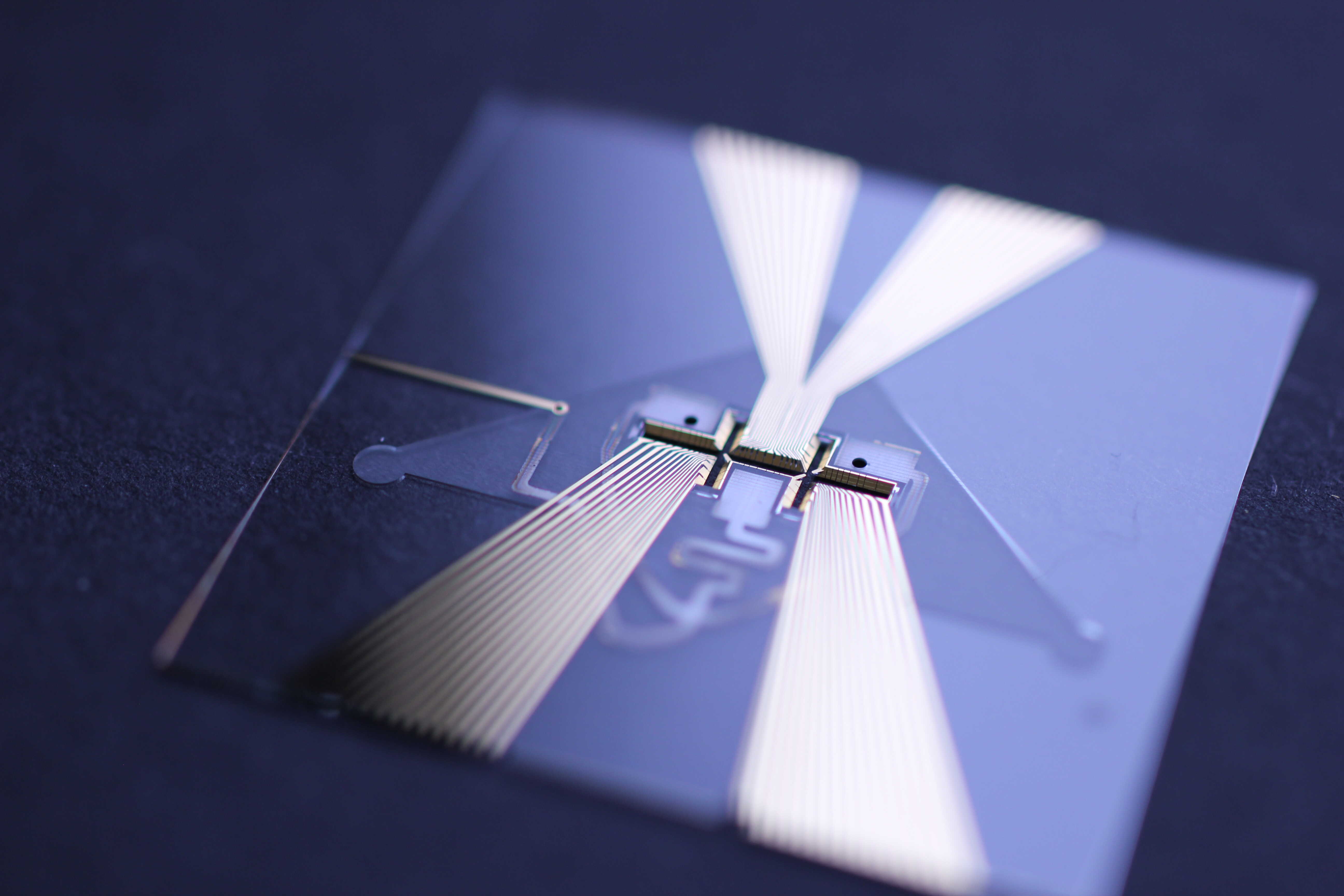}
  \put(-507,155){\crule[white]{.50cm}{.4cm}}
  \put(-505,157){\footnotesize{\textcolor{black}{\textbf{a.)}}}}
  \put(-251,155){\crule[white]{.50cm}{.4cm}}
  \put(-249,157){\footnotesize{\textcolor{black}{\textbf{b.)}}}}
\end{subfigure}
\caption{\textbf{a.)} Photograph of one electrode wafer from the cavity integrated trap after fabrication process with metal deposition of up to $5\ \mu\mathrm{m}$ thickness, 80 DC-lines (40 per wafer) and two large ground planes. \textbf{b.)} Photograph of a double junction trap wafer after evaporation of 200 nm of titanium and 700 nm of gold. The smallest electrode is 30 $\mu$m wide.}
\label{fig:TrapWafers}
\end{figure}

Once the substrates have been fabricated using laser enhanced etching, we use five standard steps to pattern electrodes and wires. All these steps are carried out in a cleanroom environment. The first step is a thorough cleaning of the wafers in a Piranha acid solution. Next the wafers are coated using electron beam evaporation. We apply a titanium adhesion layer with thicknesses on the order of 100 nm, followed by 200 nm of gold. For each wafer, we perform several evaporation steps at different angles to ensure that all desired faces are covered and use laser-cut molybdenum shadow masks to define our electrode tracks. The evaporation is reliable and produces smooth surfaces. Photographs of single wafers from each of the two designs described above are shown in Figure \ref{fig:TrapWafers}. If desired, the evaporation step is followed by electroplating, which  thickens the gold layer from a few hundred nanometers to a few microns. Once all trap wafers are coated, we proceed to assemble and to glue them. For the gluing we have used the UV-cured Epotek OG198-55 for traps which we aim to use at room-temperature, and Stycast for traps designed for cryogenic set-ups.

\onecolumngrid

\twocolumngrid

\section{Conclusions}\label{Sec:Conclusions}

We have described how laser-enhanced etching can be used to create precisely aligned multi-wafer stacks with novel features for 3-dimensional ion traps. We outlined a number of possibilities which we have investigated utilizing these techniques. We show that mechanical self-alignment structures are capable of producing alignment tolerances of below  2~$\mu$m. Based on our experience we think that these fabrication methods are well suited for small-scale multi-wafer and monolithic compact trap designs which could be used in a number of areas, from quantum information processing to frequency standards and precision metrology.

\section{Acknowledgments}\label{Sec:Conclusions}
We thank Andrea Lovera at Femtoprint for many useful discussions, and Matt Grau for comments on the manuscript. We acknowledge funding from the Swiss National Fund under grant numbers $200020\_165555$ and $200020\_179147$, and from the EU Quantum Flagship H2020-FETFLAG-2018-03 under Grant Agreement no. 820495 AQTION.

\bibliography{main}

\onecolumngrid
\newpage

\begin{figure}[H]
\begin{subfigure}{\textwidth}
\begin{subfigure}{0.5\textwidth}
    \includegraphics[width=\textwidth]{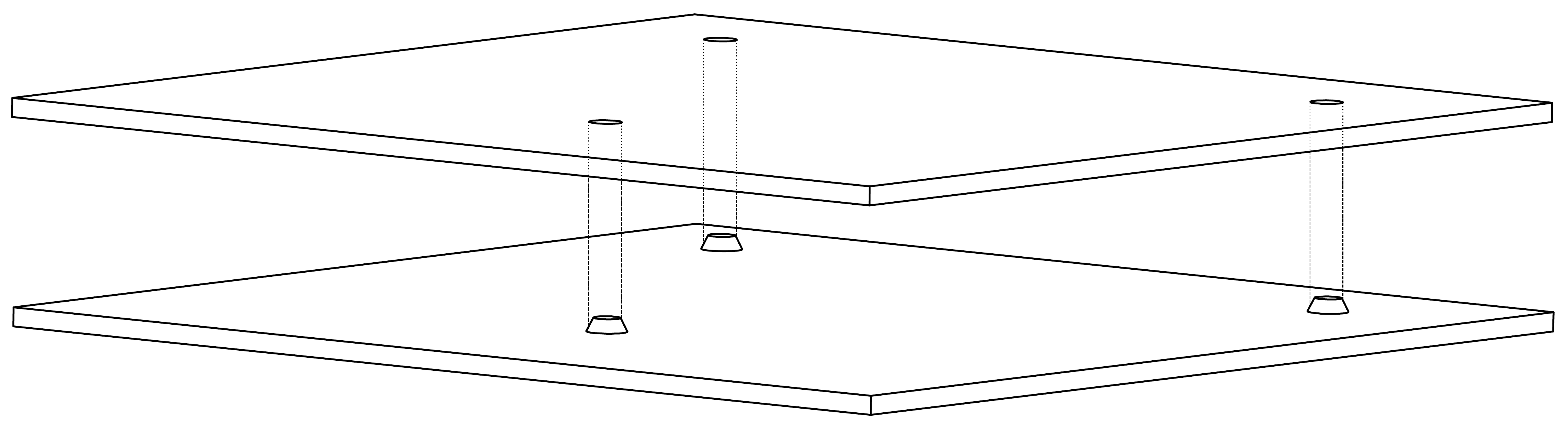}
    \put(-255,60){\footnotesize{\textcolor{black}{\textbf{a.)}}}}\\
    \includegraphics[width=\textwidth]{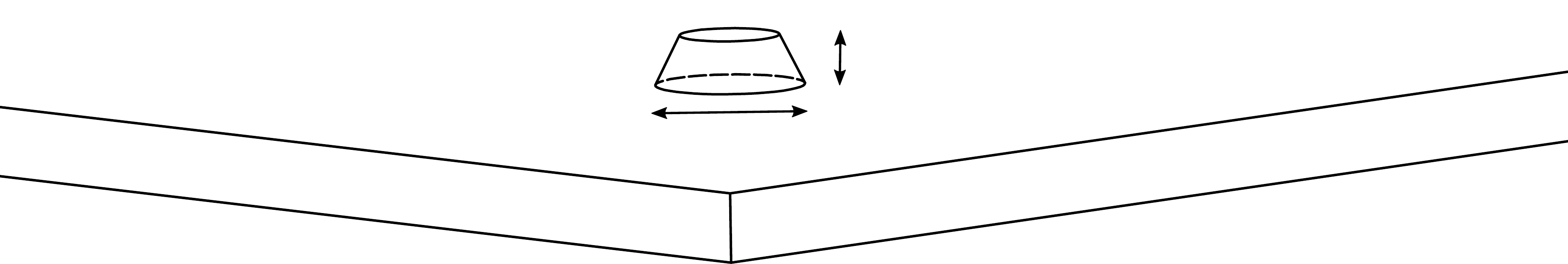}
    \put(-147,17){\tiny{\textcolor{black}{750 $\mathrm{\mu m}$}}}
    \put(-115,32){\tiny{\textcolor{black}{250 $\mathrm{\mu m}$}}}
    \put(-255,35){\footnotesize{\textcolor{black}{\textbf{b.)}}}}\\ 
    \includegraphics[width=\textwidth]{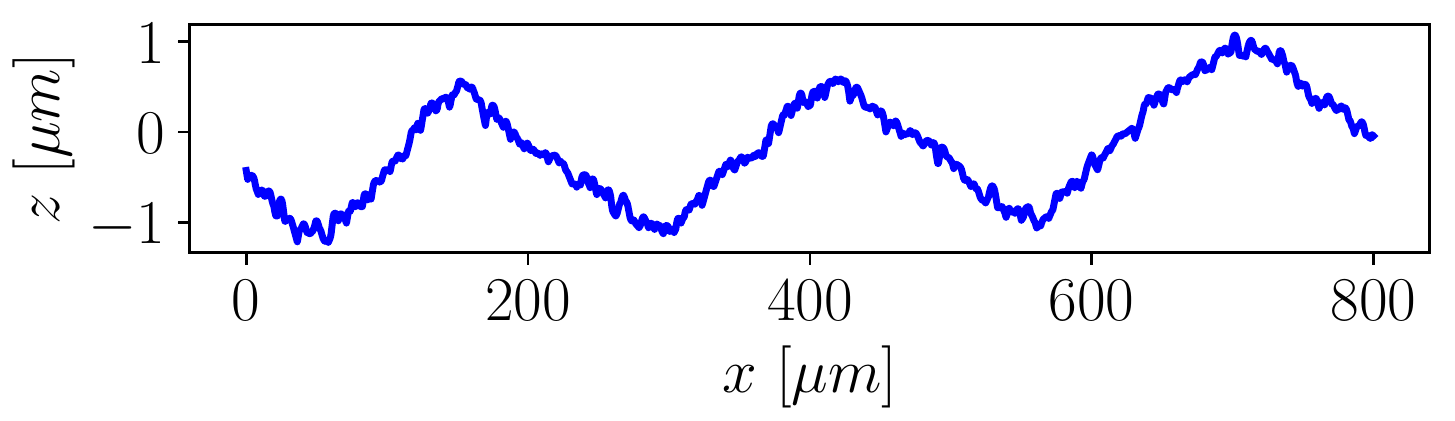}
    \put(-255,70){\footnotesize{\textcolor{black}{\textbf{d.)}}}}
\end{subfigure}
\begin{subfigure}{0.5\textwidth}
    \includegraphics[width=\textwidth]{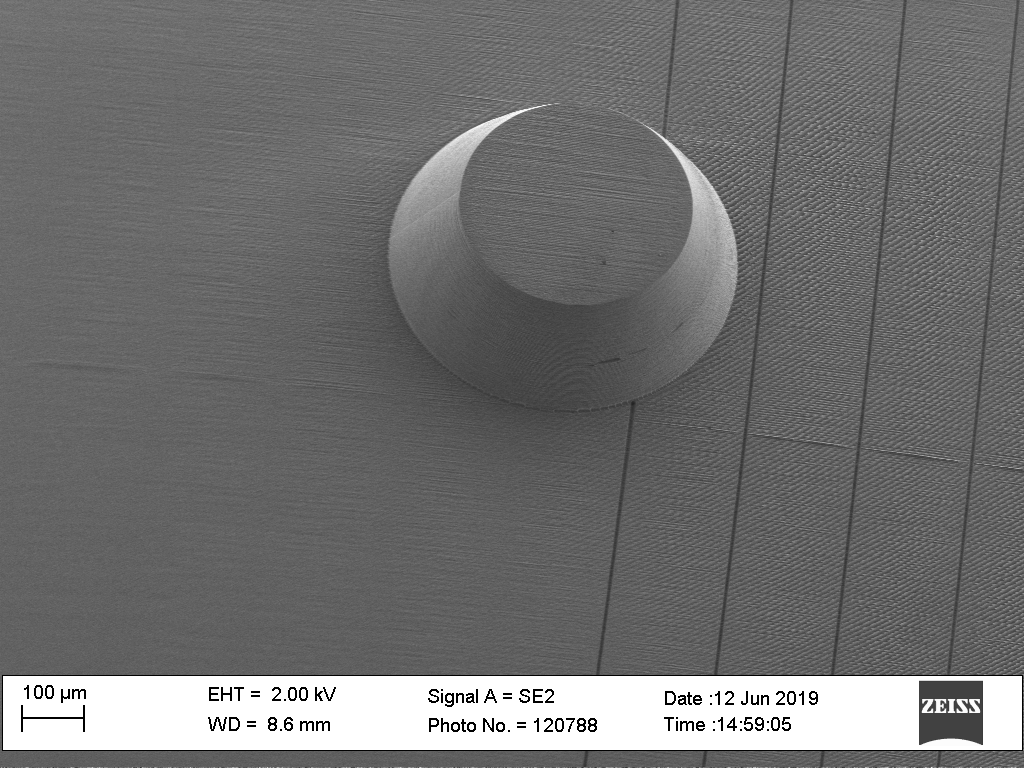}
    \put(-250,175){\crule[white]{.50cm}{.4cm}}
    \put(-248,178){\footnotesize{\textcolor{black}{\textbf{c.)}}}}
\end{subfigure}
\end{subfigure}
    \caption{\textbf{a.)} Sketch of the initial alignment strategy (exploded assembly) with alignment features (cones) integrated into one of the wafers. The initial approach was already based on three point support and the trap would have been located in the center of the wafers. \textbf{b.)} Detailed drawing of one of the alignment features, all material around it  has to be removed using femtosecond laser enhanced etching. \textbf{c.)} SEM image of one of the alignment cones and the surrounding surface. Due to the intense machining the surface quality is not very good, furthermore above mentioned defects like grooves already appear in this early sample. \textbf{c.)} Surface profile across the machining grooves.}
    \label{fig:AppFirstAlignmentTest}
\end{figure}
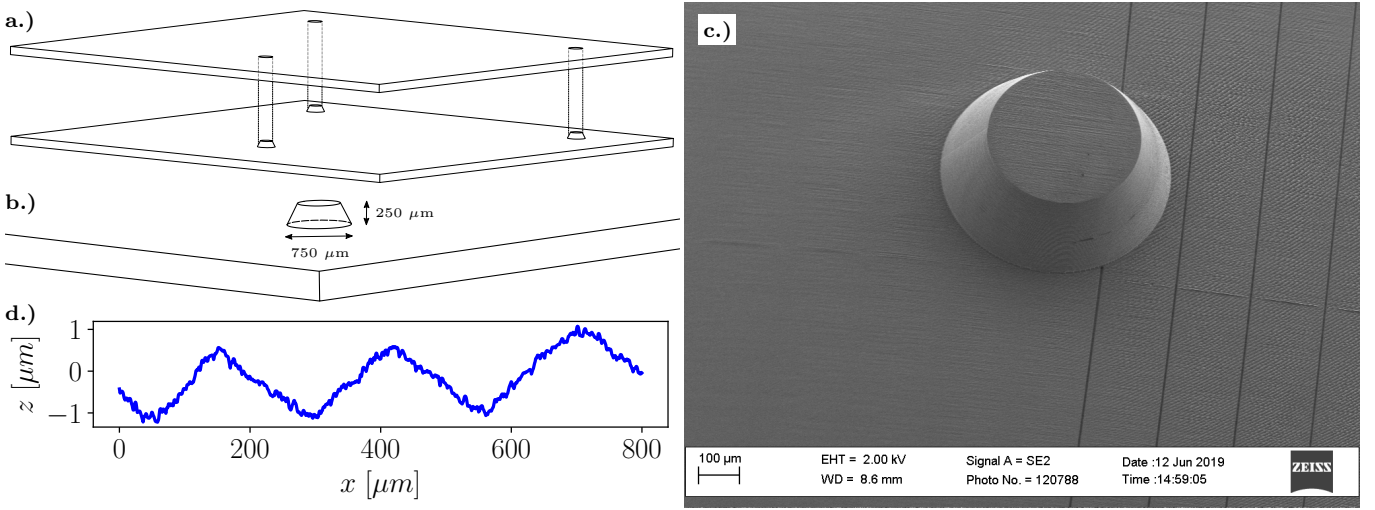
\twocolumngrid

\appendix

\section{Previous alignment approach}\label{App:WaferAlign}

The initial alignment strategy was based on two wafers with alignment features integrated into one of the wafers as illustrated in Figure \ref{fig:AppFirstAlignmentTest} \textbf{a.)}. The alignment features were small cones rising from the wafers top plane, which means that the material around them and over the full surface ($20\times 20\ \mathrm{mm}^2$) of the wafer has to be removed by femtosecond laser machining. Panel \textbf{b.)} in Figure \ref{fig:AppFirstAlignmentTest} shows a detailed drawing of one of this cones. Due to the removal of a large quantity of material, the thickness of the wafer carrying the alignment features varies by more than $20\ \mathrm{\mu m}$ and the surface roughness exceeds the one described in the main text (section \ref{Sec:LimitationsMachiningProcess}) by a factor of 10 (see Figure \ref{fig:AppFirstAlignmentTest} \textbf{d.)}). Due to stress that appeared during laser engraving the machined surface also showed the characteristic grooves described in section \ref{Sec:LimitationsMachiningProcess} as one can see in Figure \ref{fig:AppFirstAlignmentTest} \textbf{c.)}. Furthermore, we have observed a bending of the wafer which led to a variation of the distance between stacked wafers of more than $20\ \mathrm{\mu m}$.

\end{document}